\documentclass[modern]{aastex61}

\usepackage{multirow}
\usepackage[normalem]{ulem}
\usepackage{color}
\usepackage{graphicx}
\usepackage{bmpsize}

\graphicspath{{./}{images-new/}}

\received{\today}
\revised{}
\accepted{}
\submitjournal{ApJ}

\shorttitle{Coronal HXR Sources}
\shortauthors{Dennis et al.}

\begin{document}

\title{Coronal Hard X-ray Sources Revisited}

\correspondingauthor{Brian Dennis}
\email{brian.r.dennis@nasa.gov}

\author{Brian R. Dennis}
\affil{Solar Physics Laboratory, Code 671, 
NASA Goddard Space Flight Center, 
Greenbelt, MD 20771, USA}

\author{Miguel A. Duval-Poo}
\author{Michele Piana}
\affil{Dipartimento di Matematica,
Universit\`{a} degli Studi di Genova,
Genoa, 16146, Italy}

\author{Andrew R. Inglis}
\affil{Solar Physics Laboratory, Code 671,
NASA Goddard Space Flight Center,
Greenbelt, MD 20771, USA}

\author{A. Gordon Emslie}
\affil{Department of Physics \& Astronomy,
Western Kentucky University, Bowling Green, KY 42101}

\author{Jingnan Guo}
\affil{Institute of Experimental and Applied Physics, 
Christian-Albrechts-University, 24118, Kiel, Germany}

\author{Yan Xu}
\affil{Space Weather Research Lab., 
New Jersey Institute of Technology, Newark, NJ 07102}







\begin{abstract}
This paper reports on the re-analysis of solar flares in which the hard X-rays (HXRs) come predominantly from the corona rather than from the more usual chromospheric footpoints. All of the 26 previously analyzed event time intervals, over 13 flares, are re-examined for consistency with a flare model in which electrons are accelerated near the top of a magnetic loop that has a sufficiently high density to stop most of the electrons by Coulomb collisions before they can reach the footpoints. Of particular importance in the previous analysis was the finding that the length of the coronal HXR source increased with energy in the 20 - 30 keV range. However, after allowing for the possibility that footpoint emission at the higher energies affects the inferred  length of the coronal HXR source, and using analysis techniques that suppress the possible influence of such footpoint emission, we conclude that there is no longer evidence that the length of the HXR coronal sources increase with increasing energy. \textbf{In fact, for the 6 flares and 12 time intervals that satisfied our selection criteria, the loop lengths decreased on average by $1.0\pm 0.2$ arcsec between 20 and 30 keV, with a standard deviation of 3.5 arcsec.} We find strong evidence that the peak of the coronal HXR source increases in {\it altitude} with increasing energy. For the thermal component of the emission, this is consistent with the standard CHSKP flare model in which magnetic reconnection in a coronal current sheet results in new hot loops being formed at progressively higher altitudes. The explanation for the nonthermal emission is not so clear.

\end{abstract}

\keywords{Sun:activity; Sun:flares; Sun: X-rays, gamma rays}



\section{Introduction} \label{sec:intro}

  It is generally accepted that the energy release that powers a solar flare takes place in the corona. Therefore, it is of particular interest that observations from the Ramaty High Energy Solar Spectroscopic Imager \citep[RHESSI;][]{2002SoPh..210....3L} have revealed a number of flares that are characterized by a predominance of hard X-ray (HXR) emission above $\sim$20~keV from coronal sources rather than from the usually dominant chromospheric footpoint sources. For events near the limb, such as those reported by \citet{2008ApJ...678L..63K,2010ApJ...714.1108K}, this is most likely due to the occultation of the chromospheric HXR footpoints. However, this cannot explain the predominantly coronal HXR events observed on the solar disk, as discussed by \citet{2004ApJ...603L.117V}, \citet{2008ApJ...673..576X}, \citet{2011ApJ...730L..22K}, and
\citet{Guo2012,Guo2012a,Guo2013},
with only a dozen or so identified out of the RHESSI catalog of over 100,000 flares. Nevertheless, because of their relatively simple morphology and obvious connection to the primary energy release region, they are of considerable interest.  They provide the closest and most direct window currently available into the physical properties of the site(s) of flare energy release, and of the processes that govern electron acceleration and propagation in solar flares. 
  
In a series of papers over the last decade, various authors have explored the morphology of such coronal hard X-ray sources, particularly the variation of source size with energy \citep{2008ApJ...673..576X,Guo2012,Guo2012a,Guo2013,2011ApJ...730L..22K}, with the intriguing result that the extent of these coronal sources ($L$) generally increases with photon energy ($\mathbf{\epsilon}$)  at energies between $\sim$20~and~$\sim$30~keV. Such behavior is inconsistent with a thermal source, the size of which generally decreases with increasing energy as the emission becomes more and more dominated by the hottest regions \citep{2008ApJ...673..576X}, but it is consistent with the transport of accelerated electrons through a collisional target, since higher energy electrons travel further.
  
The {\em RHESSI} data encodes spatial information about the X-ray source structure in a set of spatial Fourier components, commonly called ``visibilities.'' The source centroid can be located relatively easily from data in this form since it depends  straightforwardly on the modulation frequency and phasing of each rotating modulation collimator (RMC). However, the source dimensions -- basically the length and width -- are more difficult to determine since they depend on the relative amplitudes of the modulation in each of the RMCs that have angular resolutions comparable to the appropriate source dimension. 

Previous analyses used visibilities derived from the count rates in the various {\em RHESSI} detectors to construct images of the X-ray photon flux. \citet{Guo2012a} also spectrally inverted the observed count visibilities to obtain electron visibilities \citep{2007ApJ...665..846P}, which were then used to construct images of the mean electron flux \citep{2003ApJ...595L.115B}.
From these photon and electron images, the variation of $L$ with photon energy $\epsilon$ and electron energy $E$ was determined for a number of events.  The form of this variation, interpreted as the increase in propagation distance with electron energy in a cold thick target, can then be used to determine both the length of the acceleration region and the density of the medium in which the accelerated electrons propagate \citep{Guo2012a}. In all cases studied, the column depth was found to be sufficiently high to stop the bulk of the accelerated electrons in the coronal part of the flare loop. As shown by \citet{2008ApJ...673..576X} and \citet{Guo2012,Guo2012a,Guo2013}, such thick-target coronal sources can provide us with substantial information on the distribution of accelerated electrons, from their initial acceleration out of the background thermal distribution to their ultimate re-thermalization.

In the analyses above, it was simply assumed that the extended sources were wholly coronal and that they extended along the magnetic field direction in a single confined loop.  Further, the field-aligned extent of the source was generally estimated from integral moments of the observed flux, either the ``one-sided first-order moment'' \citep{2008ApJ...673..576X} or the second-order moment \citet{Guo2012,Guo2012a}.  However, these assumptions and techniques are suspect, for the following reasons:

\begin{enumerate}

\item The use of integral moments to determine the ``length'' of a coronal source means that any chromospheric footpoint emission present in the image, even at low intensity levels, can have a significant impact. (This is especially true if a second-order integral moment is used). Further, it is likely that the footpoint sources will have a significantly harder spectrum than the coronal source, since the latter will include some thermal emission at lower energies. Thus, even weak footpoint sources that are not clearly visible but which are situated at large distances from the source centroid can have a significant effect on the moments at higher energies, leading to an inferred increase in source size with energy. 

\item It is particularly difficult to separate coronal and footpoint emission for loops that are not viewed face on, i.e., from a direction perpendicular to the plane of the loop. If the loop is viewed from above, for example, it will be impossible to separate the emission from the legs of the loop and from the footpoints. Consequently, the most favorable location on the solar disk for the flare for making measurements of the length of the coronal part of the loop source is near either limb when the footpoints are aligned close to the north-south direction. 

\item Also of concern is that multiple magnetic loops are likely to be involved in the energy release and particle acceleration process. Hence, the HXR emission at different energies may not all be from the same location. For example, there is considerable evidence that in some coronal sources the higher energy emission originates from a different location, most likely at a higher altitude, than the lower energy emission \citep{2002SoPh..210..341G, 2003ApJ...596L.251S, 2004ApJ...612..546S, 2015A&A...584A..89J}. Variations in source altitude with time have also been reported in several cases \citep[e.g.,][]{Gallagher2002,2006A&A...446..675V}. Such variations in source altitude with energy and time were not considered in the earlier work but must be taken into account in any model that seeks to interpret the measured variation in the extent of the coronal HXR source with energy.

\end{enumerate}

The main objective of the present work is to re-examine the evidence for variation of the length of coronal HXR sources as a function of energy in light of the above concerns.  Care is taken to determine the most likely locations for any footpoint emission at the higher energies and to ensure that it is not included in the estimation of the extent of the coronal source along the magnetic loop at any energy. Also, any change in the peak and/or centroid location of the coronal source with energy is noted.  The following aspects of the current investigation are worthy of note:

\begin{enumerate}

\item Different image reconstruction algorithms are used, including MEM NJIT \citep{2007SoPh..240..241S}, EM \citep{benvenuto2013expectation}, and VIS WV \citep{duval2017compressed,2018A&A...615A..59D}.  This allows not only the values of parameters associated with the source structure to be evaluated but also their quantitative uncertainties;

\item More sophisticated techniques are used to separate coronal emission from footpoint emission in determining the inferred length of the coronal HXR source;

\item More detailed spectral analyses are carried out to better evaluate the thermal and nonthermal components of the X-ray emission as a function of energy for each event.

\item Observations from other instruments, such as the Transition Region and Coronal Explorer 
(\citealp[TRACE,][]{1999SoPh..187..229H}) and the Atmospheric Imaging Assembly (\citealp[AIA,][]{2012SoPh..275...17L}) on the Solar Dynamics Observatory (\citealp[SDO,][]{2012SoPh..275....3P}), are used to place the hard X-ray images into the context of the overall flare morphology. Since all but three of the original coronal HXR source events previously analyzed by \citet{2004ApJ...603L.117V}, \citet{2008ApJ...673..576X}, and \citet{Guo2012,Guo2012a,Guo2013} occurred before SDO was launched, we have used other imaging information, notably from TRACE, to help with determining the possible location of footpoints and the general magnetic field topology in the flaring region.

\end{enumerate}

In Section~\ref{section-DataAnalysis}, we present the new method used to analyze the size, shape, and location of coronal HXR sources.  In Section~\ref{summary-of-parameters} we present a summary of the results for those events deemed to have properties that can be reliably determined at a statistically significant level. In Section~\ref{conclusions} we present our conclusions and their physical significance.

\section{Data Analysis}
	\label{section-DataAnalysis}
    
We have reanalyzed all 13 flares (some with multiple time intervals) studied by \citet{2004ApJ...603L.117V, 2008ApJ...673..576X}, and \citet{Guo2012, Guo2012a, Guo2013}, plus one additional event on 15 May 2013 that was observed with both RHESSI and AIA.  The full list of events is given in Table~\ref{tab-parameters}, and a summary of the issues in determining reliable source lengths is given in Appendix~\ref{Appendix}. In each time interval, we have assessed the extent to which the original single dense loop model is consistent with the observations, and, where appropriate, suggested alternative geometries. We have used different image reconstruction techniques and different analysis methods to estimate the length of the source along the toroidal direction of a loop with the expected orientation as seen for the specific location on the solar disk.  We have been mindful of the possible influence of footpoint emission, particularly at higher energies, on the measured source length. It is not always clear from the RHESSI images alone where the footpoints would be. Hence, where possible, we have used UV/EUV images to locate the flare ribbons and hence the likely location of the loop footpoints.

\begin{table}[!ht]
		\caption{Dates, times, and locations for all 14 analyzed flares.}
		\label{tab-parameters}
		\centering		
		\begin{tabular}{|r|c|c|c|r|r|c|c|c|c|c|}
			\hline\hline
			\multirow{4}{*}{\#} & \multirow{4}{*}{Date} & \multirow{4}{*}{Time (UT)} &  &
            \multicolumn{2}{c|}{Peak} & \multicolumn{4}{c|}{Footpoints} & \multirow{4}{*}{Y/N} \\
            
			\ &  &  & GOES & \multicolumn{2}{c|}{(arcsec)} & \multicolumn{4}{c|}{(arcsec)} & \\
            
			\cline{5-10}
			\ & & & Class & \multirow{2}{*}{X~} & \multirow{2}{*}{Y~} & \multicolumn{2}{c|}{East} & \multicolumn{2}{c|}{West} & \\
            
			\cline{7-10}
			\ & \ & \ & \  & \ & \ & X & Y & X & Y & \\
			\hline		
			\hline
			1 & \multirow{2}{*}{12-Apr-2002} & 17:42:00--17:44:32 & \multirow{2}{*}{M4.1} & 408 & 448 & --~ & --~ & --~ & --~ & N \\
			2 &             & 17:45:32--17:48:32 & & 415 & 446 & --~ & --~ & --~ & --~ & N \\
			\hline
			3 & \multirow{3}{*}{15-Apr-2002} & 00:00:00--00:05:00 & \multirow{3}{*}{M3.7} & 781 & 382 & 760 &  390 &  770 & 370 & Y \\
			4 &             & 00:05:00--00:10:00 & & 783 & 383 &  ~   & ~ & ~ &  ~ & Y \\
			5 &             & 00:10:00--00:15:00 & & 789 & 379 &~&~&~&~& Y \\
			\hline
			6 & \multirow{2}{*}{15-Apr-2002} & 23:05:00--23:10:00 & \multirow{2}{*}{M1.2} & 877 & 359 & 845 & 370 & 863 & 350 & Y \\
			7 &             & 23:10:00--23:15:00 & & 877 & 356  &~&~&~&~&  Y \\		
			\hline
			8 & \multirow{2}{*}{17-Apr-2002} & 16:54:00--16:56:00 &  \multirow{2}{*}{C9.8} & 927 & -245 & -- & -- & -- & -- &N \\
			9 &             & 16:56:00--16:58:00 & & 928 & -246 & -- & -- & -- & -- & N \\
			\hline
			10 & \multirow{2}{*}{17-Jun-2003} & 22:46:00--22:48:00 &  \multirow{2}{*}{M6.8} & -810 & -135 & -- & -- & -- & -- & N \\
			11 &             & 22:48:00--22:50:00 & & -812 & -145& -- & -- & -- & -- & N \\
			\hline
			12 & \multirow{2}{*}{10-Jul-2003} & 14:14:00--14:16:00 &  \multirow{2}{*}{M3.6} & 940 & 216 & -- & -- & -- & -- & N \\
			13 &             & 14:16:00--14:18:00 & & 940 & 215 & -- & -- & -- & -- & N \\
			\hline
			14 & \multirow{2}{*}{21-May-2004} & 23:47:00--23:50:00 &  \multirow{2}{*}{M3} & -757 & -157 & -745 & -140 & -742 & -163 & Y \\
			15 &             & 23:50:00--23:53:00 & & -757 & -157 &~&~&~&~&  Y \\
			\hline
			16 & \multirow{3}{*}{31-Aug-2004} & 05:31:00--05:33:00 &  \multirow{3}{*}{M1.4} & 940 & 95 & -- & -- & -- & -- & N \\
			17 &             & 05:33:00--05:35:00 & & 940 & 95 & -- & -- & -- & -- & N \\
			18 &             & 05:35:00--05:37:00 & & 940 & 95 & -- & -- & -- & -- & N \\
			\hline
			19 & \multirow{2}{*}{01-Jun-2005} & 02:40:20--02:42:00 &  \multirow{2}{*}{M1.7} & -689 & -292 & -- & -- & -- & -- & N \\
			20 &             & 02:42:00--02:44:00 & & -690 & -294 & -- & -- & -- & -- & N \\
			\hline
			21 & \multirow{2}{*}{23 Aug-2005} & 14:23:00--14:27:00 &  \multirow{2}{*}{M2.7} & 925 & -240 & 890 & -200 & 880 & -240 & Y \\
			22 &             & 14:27:00--14:31:00 &  & 923 & -235  &~&~&~&~&  Y \\		
			\hline
			23 & \multirow{2}{*}{13-Feb-2011} & 17:33:00--17:34:00 &  \multirow{2}{*}{M6.6} & -77 & -233 & -- & -- & -- & -- & N \\
			24 &             & 17:34:00--17:35:00 & & -79 & -233 & -- & -- & -- & -- & N \\
			\hline
			25 & 03-Aug-2011 & 04:31:12--04:33:00 &  M1.7 & -156 & 166 & -157 & 174 & -147 & 162 & Y \\
			\hline
			26 & 25-Sep-2011 & 03:30:36--03:32:00 & C7.9 & -704 & 154 & -- & -- & -- & -- & N\\
			\hline
			27 & \multirow{2}{*}{15-May-2013} & 01:37:40--01:38:28 &  \multirow{2}{*}{X1.2} & -881 & 194 & -850 & 178 & -845 & 200 & Y \\
			28 &             & 01:38:28--01:39:44 & & -881 & 196  &~&~&~&~&  Y \\
			\hline\hline
		\end{tabular}		
        \begin{description}		
			\item [\#, Date, and Time] Intervals used in chronological order.
			\item [Peak and Footpoints] X and Y locations of the flare peak flux and footpoints.
			\item [Y/N] Indicates if results for this time interval were/were not used for subsequent analysis.
		\end{description}
\end{table}

In the following section, we illustrate our new method of analysis by showing the results for one of the best examples of these HXR coronal sources, the M-class flare on 14/15 April 2002 that was also analyzed by \citet{2004ApJ...603L.117V}, \citet{2004ApJ...612..546S}, and \citet{2011ApJ...730L..22K}.  We describe the procedures used to critically assess the observations and to determine the extent to which they support the original single dense loop model. In later sections, we present a summary of the results of our analysis for all of the events.

\subsection{Flare on 14/15 April 2002}
	\label{sub-15April2002}

\begin{figure}
   \includegraphics*[width=1.0\textwidth, angle = 0, 
   trim = 0 65 0 0]
   		{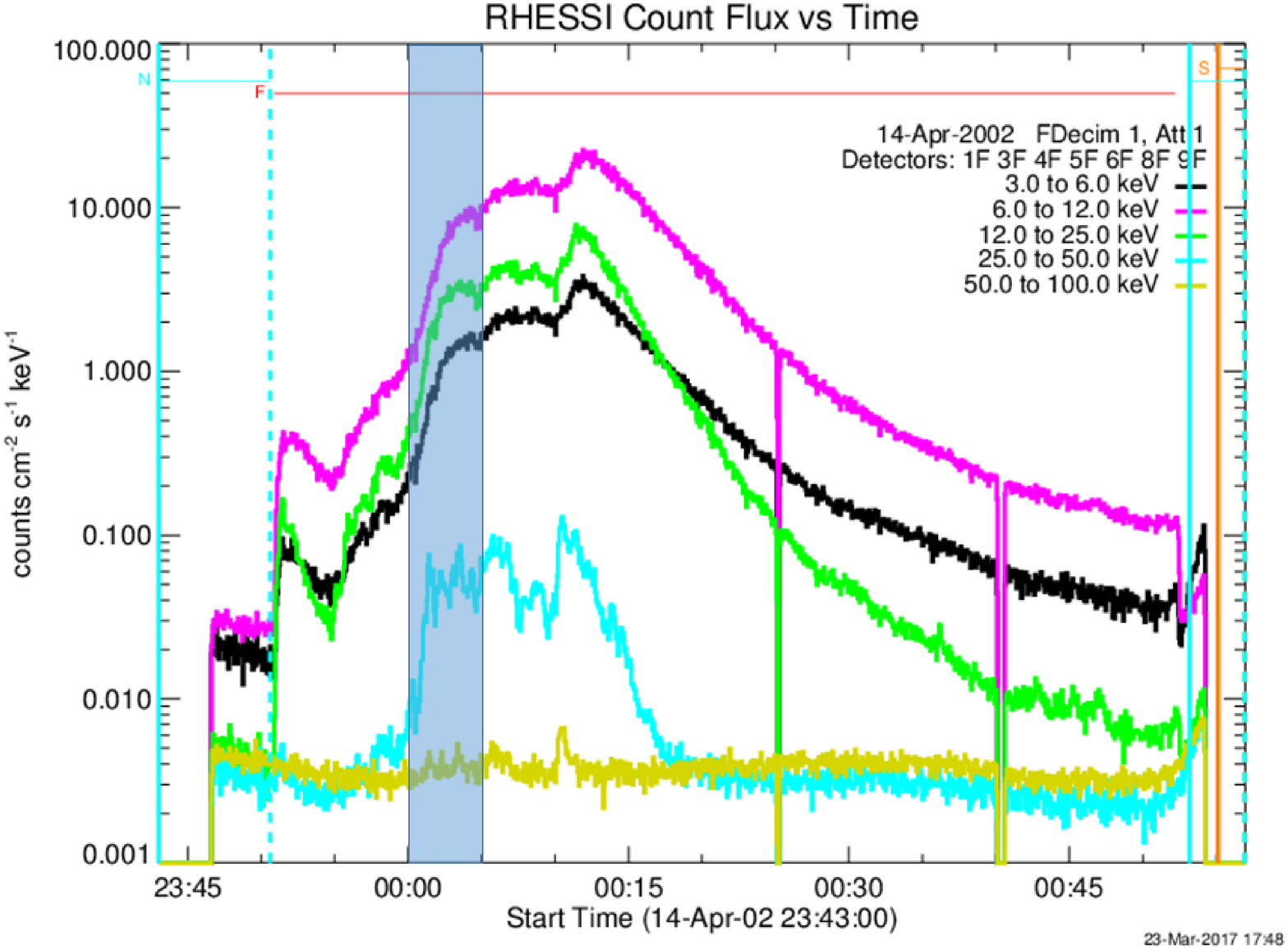}
    \caption{RHESSI light curves for the flare on 15 April 2002. The color-coded curves are for the five indicated energy ranges. Counts from the front segments of all detectors except for detectors \#2 and \#7 were summed and divided by the summed live times and the total effective sensitive area of 35.59~cm$^2$ per detector to give the plotted values with a 4~s cadence to match the spacecraft spin period. The thin attenuators were in place above all detectors limiting the useful energy range to $>6$~keV. The blue shaded areas show the first of the three time intervals between 00:00 and 00:05 UT used here and by \citet{Guo2012,Guo2012a,Guo2013} to determine the source dimensions.}
   \label{Fig-lc15April2002}
   \end{figure}

This M3.7-class flare occurred at N19W60 in NOAA active region 09893; the GOES event started at 23:34 UT on the 14th of April 2002, peaked at 00:14 UT on the 15th, and ended at 00:25 UT.  The RHESSI time profiles in five broad energy bins are shown in Figure~\ref{Fig-lc15April2002} with the first time interval used here and by \citet{Guo2012,Guo2012a,Guo2013} shown by the blue box.

\begin{figure}
    \includegraphics*[width=0.8\textwidth, angle = 90, 
   trim = 0 0 0 0]
{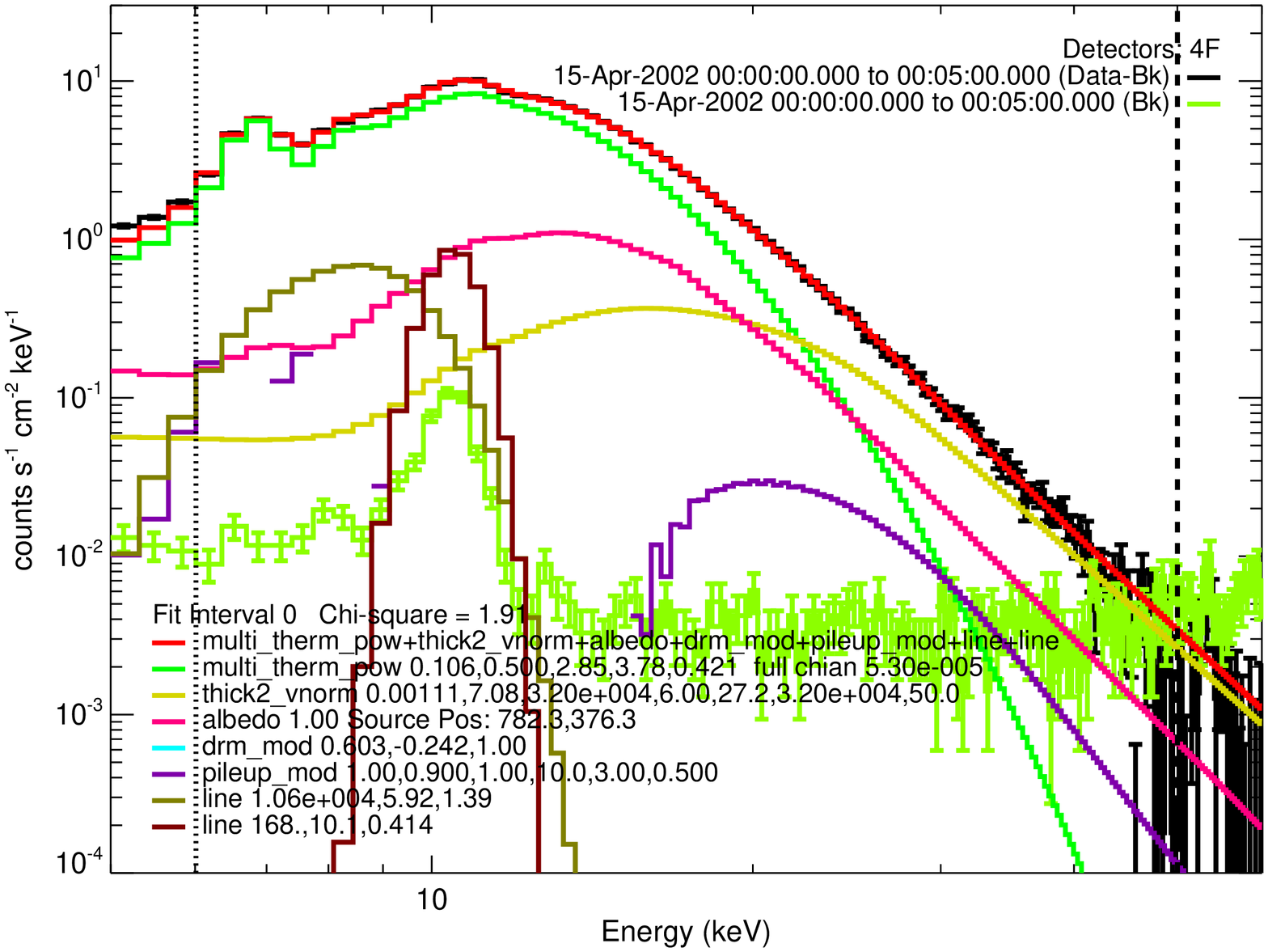}
    \caption{RHESSI count flux spectrum for the five-minute time interval shown in \mbox{Figure~\ref{Fig-lc15April2002}.} The black histogram with $\pm 1 \sigma$ statistical error bars is the background-subtracted count flux in the front segment of Detector \#4. The red histogram is the function that was fitted to the data between 6 and 50 keV with a reduced $\chi^2$ value of 1.9.  It is the sum of the following components: a multi-thermal bremsstrahlung function (green), a power-law nonthermal thick-target function (yellow), an albedo function for isotropic emission (pink), the estimated pulse pile-up contribution (purple), and two Gaussian instrumental lines (olive and brown). The background spectrum (green) determined from the nighttime period immediately prior to the flare is shown with $\pm 1 \sigma$ error bars. The values of all parameters used for the fit are given for each functional component.}
   \label{Fig-sp15April2002}
   \end{figure}
   
The increases in X-ray emission above the RHESSI background counting rates are clearly seen in Figure \ref{Fig-lc15April2002} at all energies, including small increases in the 50-–100~keV range.  The measured count flux spectrum in Figure \ref{Fig-sp15April2002} is for the five-minute time interval used to make the images in Figure~\ref{Fig-lc15April2002}. It was generated by fitting the background-subtracted count-flux spectrum from the front segment of Detector \#4 with the predicted flux from an assumed photon spectrum made up of a thermal and a nonthermal component plus an albedo component and an estimated pulse-pile-up contribution. Two instrumental Gaussian lines were added to fit the data below $\sim$12~keV. Recognizing that a range of temperatures exists in the thermal plasma, the thermal component was modeled by a differential emission measure (DEM) with a power-law dependence on temperature (T):

\[DEM = (0.11\pm 0.01) \, (2/T)^{3.8\pm 0.2} \times 10^{49} \, {\rm cm}^{-3} \, {\rm keV}^{-1}; \,  \, 0.5 \le T \le (2.8 \pm 0.2) \, {\rm keV} \,\,\, .\]
The nonthermal component was modeled as the photon spectrum expected from the injection into a cold, thick target \citep{1971SoPh...18..489B} of a flux of electrons with a power-law spectrum (index of $7.1\pm 0.1$) and a sharp low-energy cutoff of $\leq27$~keV.

The fitted spectral components in Figure \ref{Fig-sp15April2002} show that the thermal and nonthermal components have equal count fluxes at $\sim$22 keV.  This is in agreement with the impulsive nature of the 25 - 50 keV light curve in Figure \ref{Fig-lc15April2002} but it is in disagreement with the transition energy of 15~keV given by \citet{Guo2012}.
The difference is because they assumed a thermal function with a single temperature of 1.6~keV instead of the extension up to 2.8~keV used here.  This is a more likely situation since an isothermal model cannot account for the changes in source altitude with energy. A multi-thermal coronal source is required with a differential emission measure extending over a significant temperature range as with the power-law, exponential, or Gaussian dependency used by \cite{2015A&A...584A..89J}. 

\subsection{Identification of source geometry}
	\label{sub-goemetry}
\begin{figure}
{\centering
    \includegraphics*[width=0.9\textwidth, angle = 90, 
   trim = 20 80 0 80]
{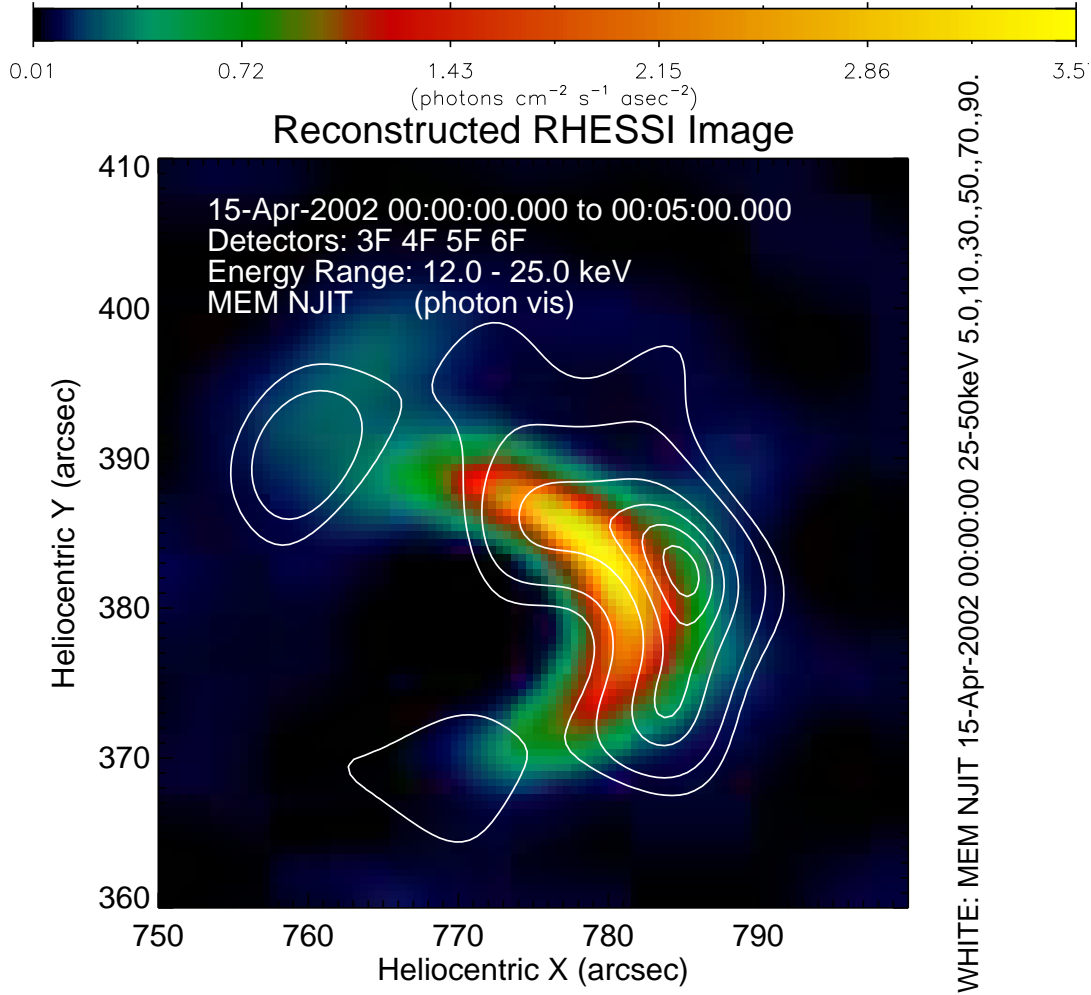}}
    \caption{RHESSI 12 - 25 keV image (color-coded according to the color bar in units of photons~cm$^{-2}$~s$^{-1}$~arcsec$^{-2}$) overlaid with 25 - 50 keV contours (5, 10, 30, 50, 70, and 90\% of the peak value) for the 5-min.~interval from 00:00 to 00:05~UT made using the MEM\_NJIT reconstruction technique with data from the front segments of detectors 3, 4, 5, and 6.  Two compact footpoints are evident in the 25 - 50 keV image with the extended coronal source present at both energies but at a higher altitude in the higher energy image.}
   \label{Fig-im15April2002-12to50kev}
   \end{figure}

In order to determine the extent of the coronal source, we first created images in different photon energy bins for the time interval from 00:00 to 00:05 UT indicated in Figure \ref{Fig-lc15April2002}, the same interval used by \citet{Guo2012,Guo2012a}. Images made in two broad energy bins are shown in Figure \ref{Fig-im15April2002-12to50kev} with the colors representing the flux in the 12 - 25 bin and the overlaid white contours representing the flux in the 25 - 50 keV bin.  Two compact footpoints are evident in the 25 - 50 keV image; their centroid locations were used for the subsequent analysis in narrower energy bins.  The extended coronal source is present at both energies but further to the west in the higher energy image,  corresponding to a higher altitude.

\begin{figure}[htp]
  \centering
  \includegraphics[width=0.24\textwidth]{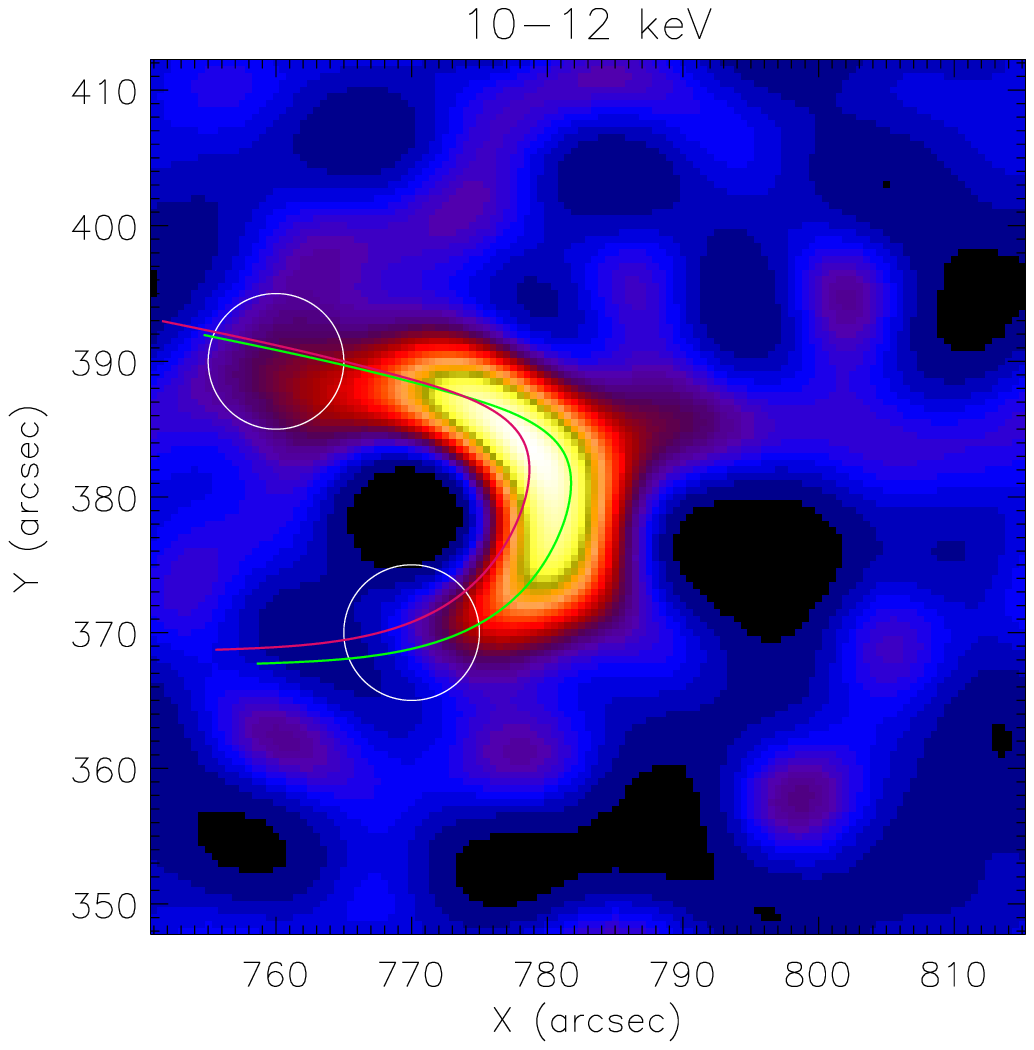}
  \includegraphics[width=0.24\textwidth]{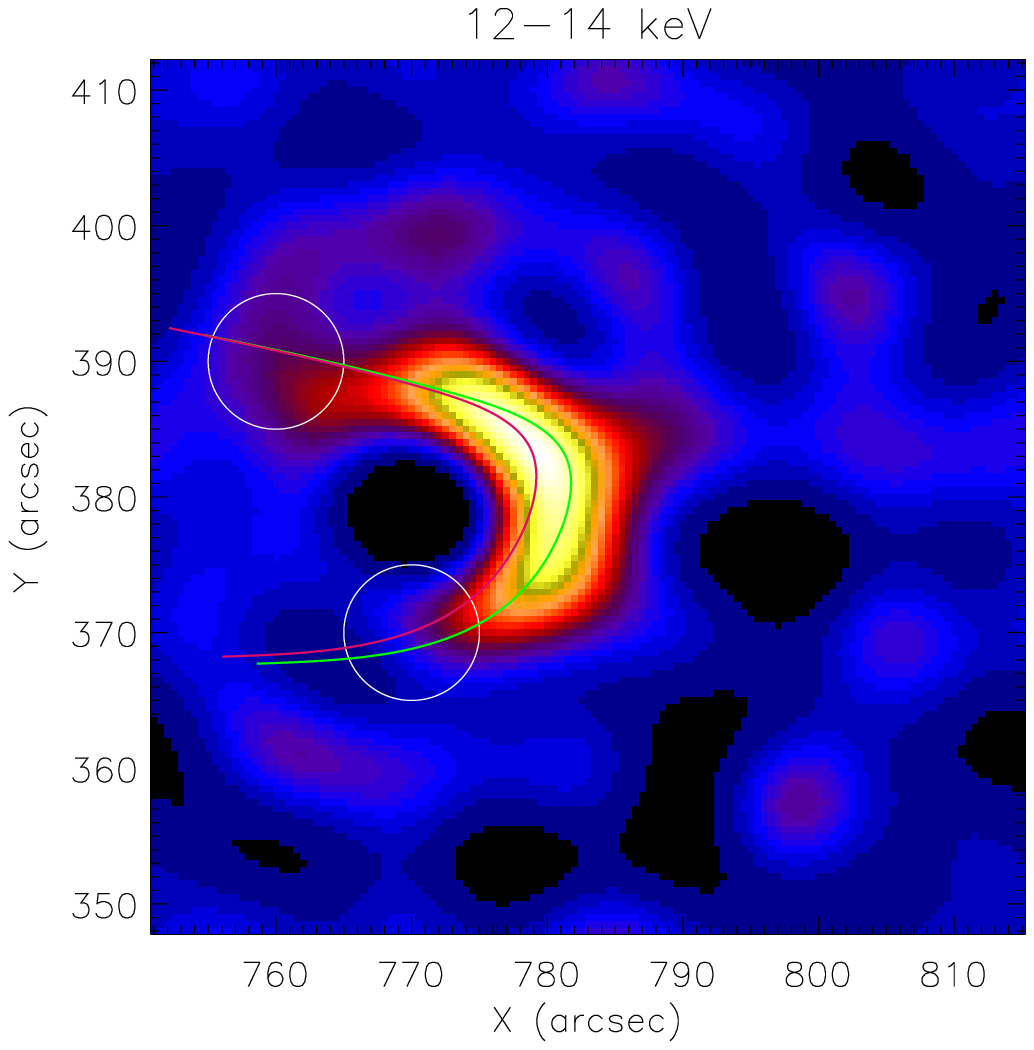}
  \includegraphics[width=0.24\textwidth]{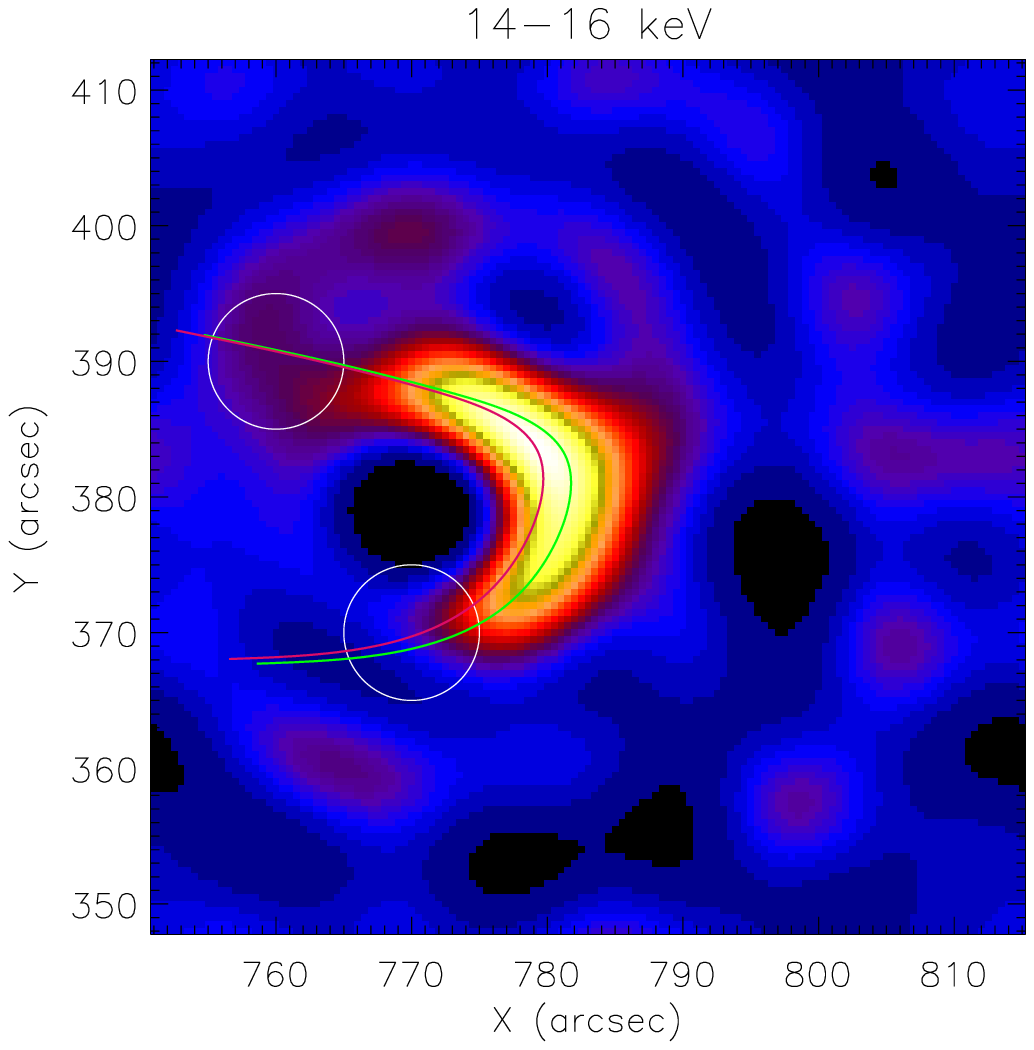}
  \includegraphics[width=0.24\textwidth]{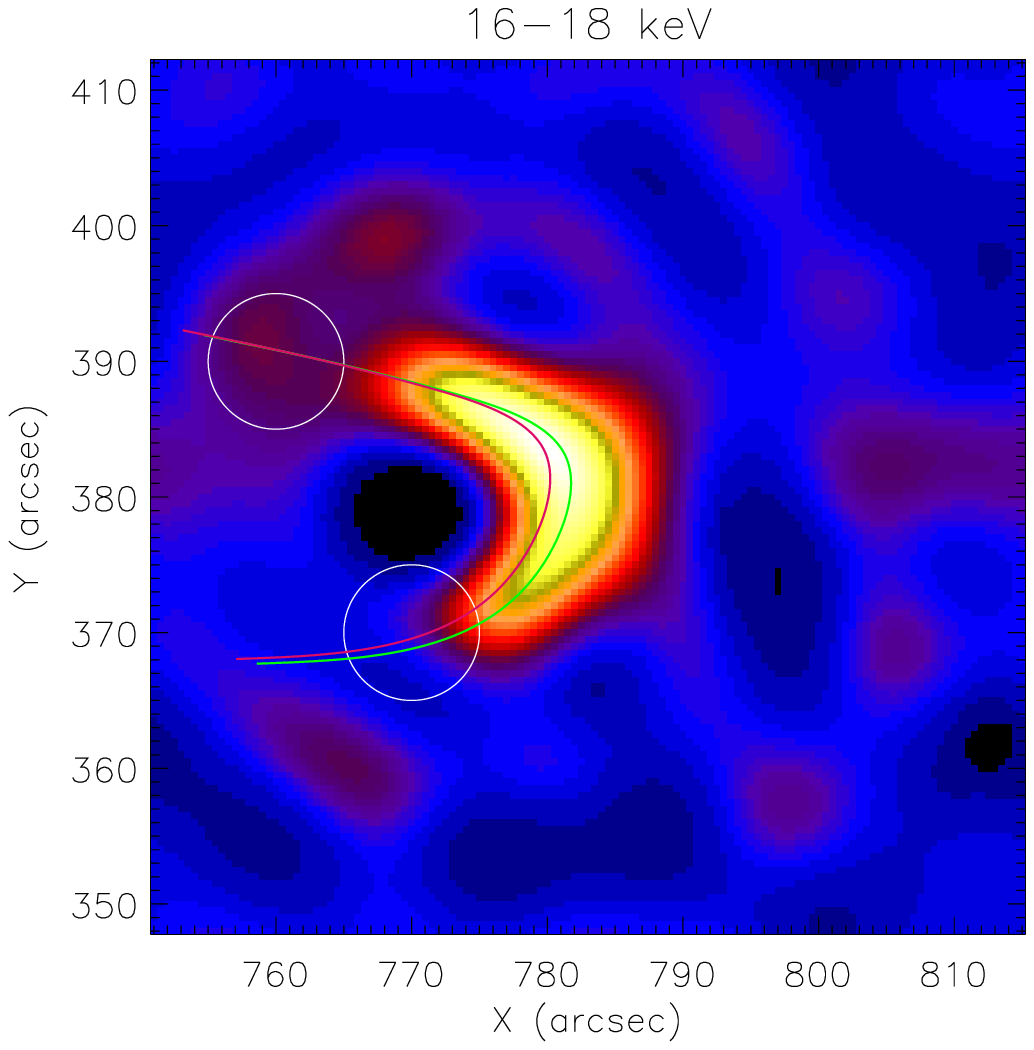}\\ 
  \includegraphics[width=0.24\textwidth]{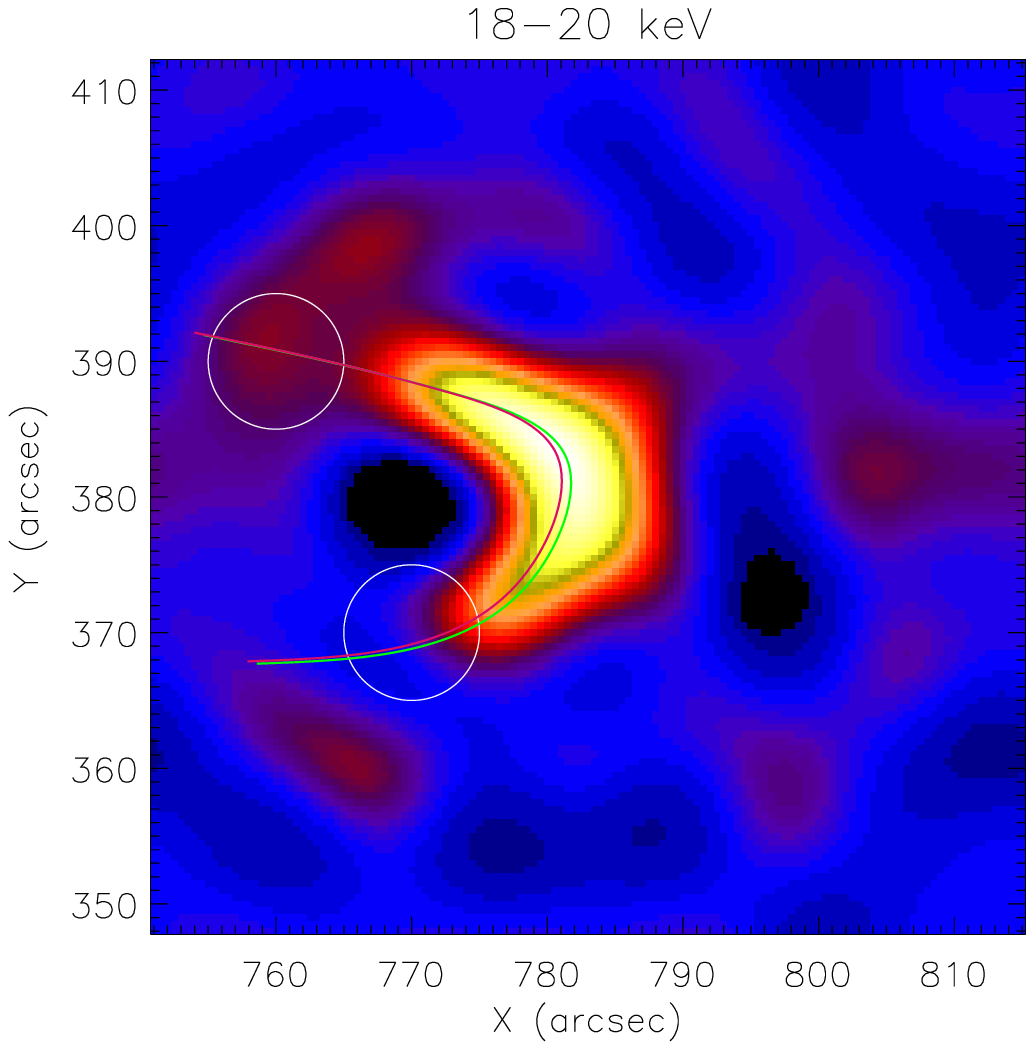}
  \includegraphics[width=0.24\textwidth]{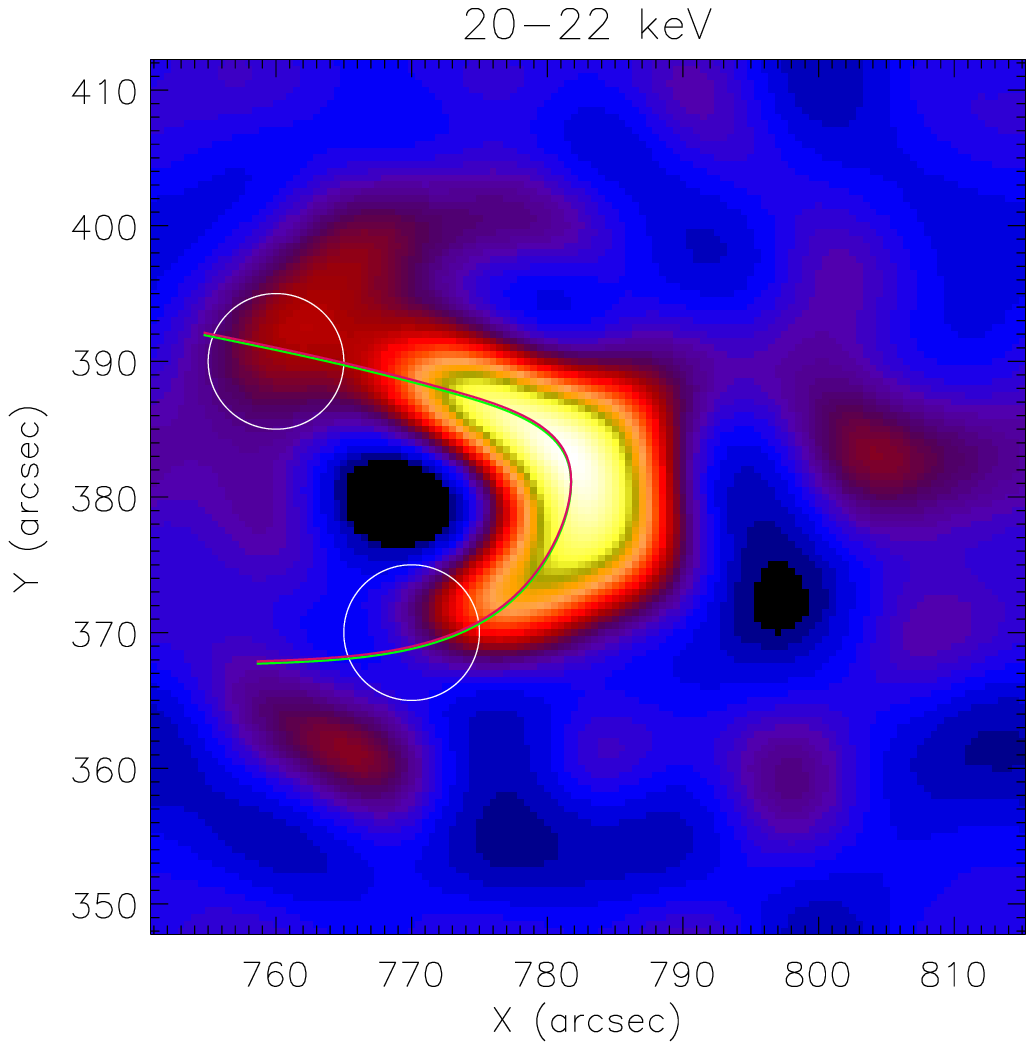}
  \includegraphics[width=0.24\textwidth]{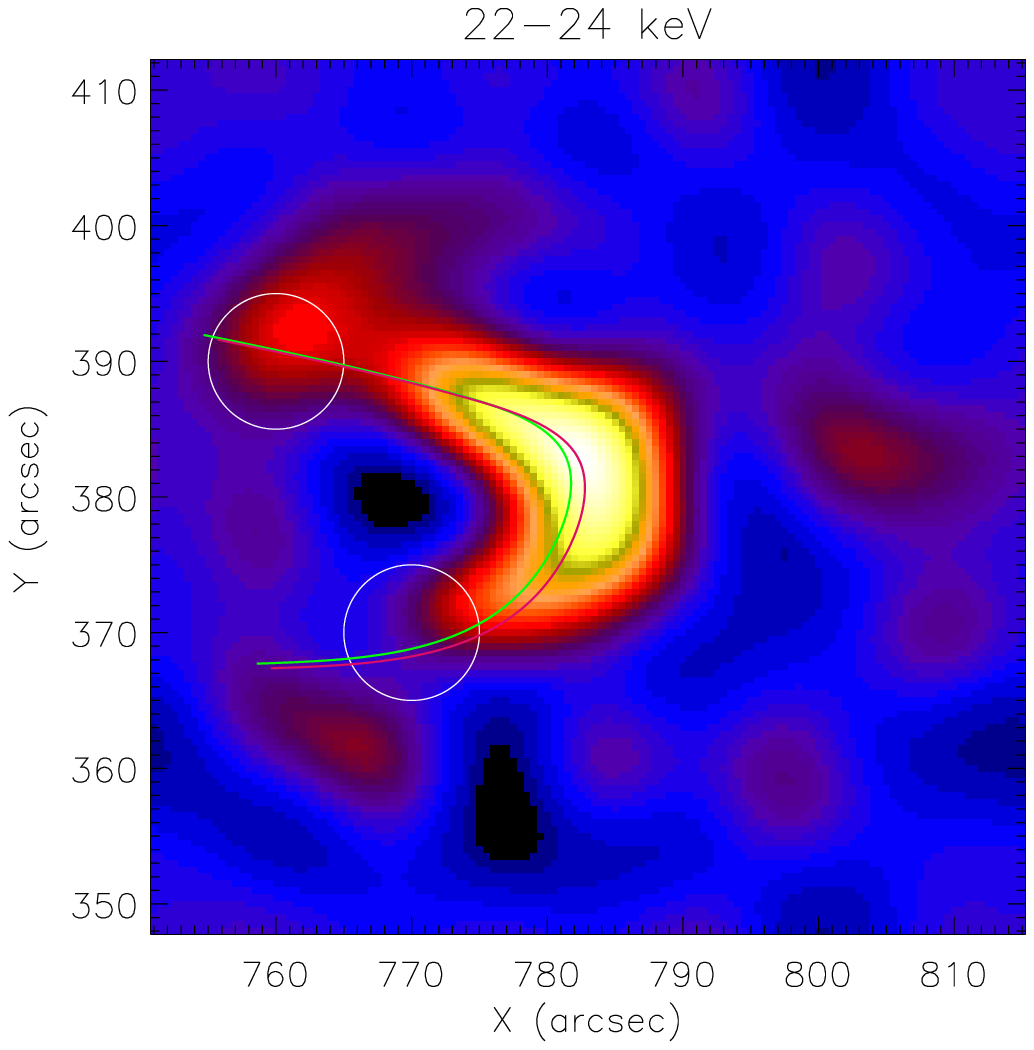}
  \includegraphics[width=0.24\textwidth]{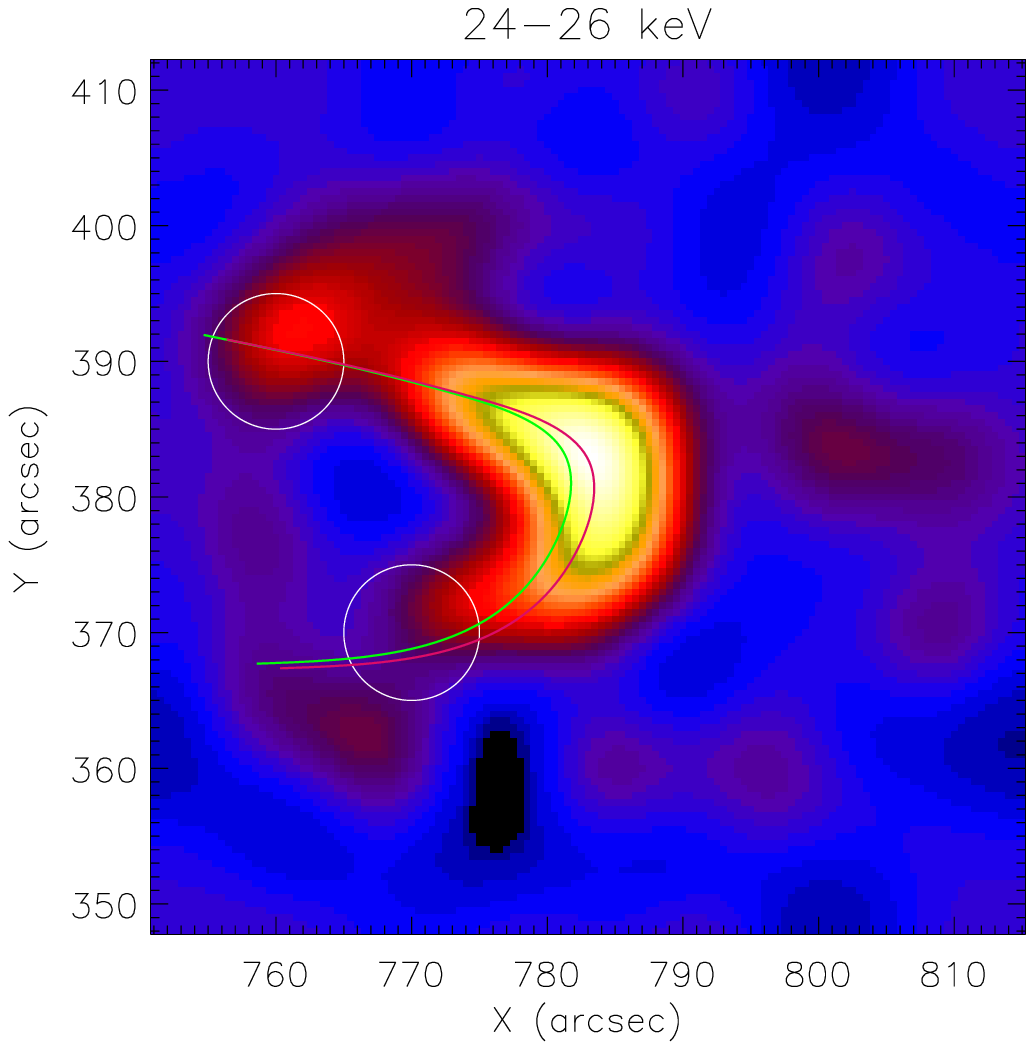}\\
  \includegraphics[width=0.24\textwidth]{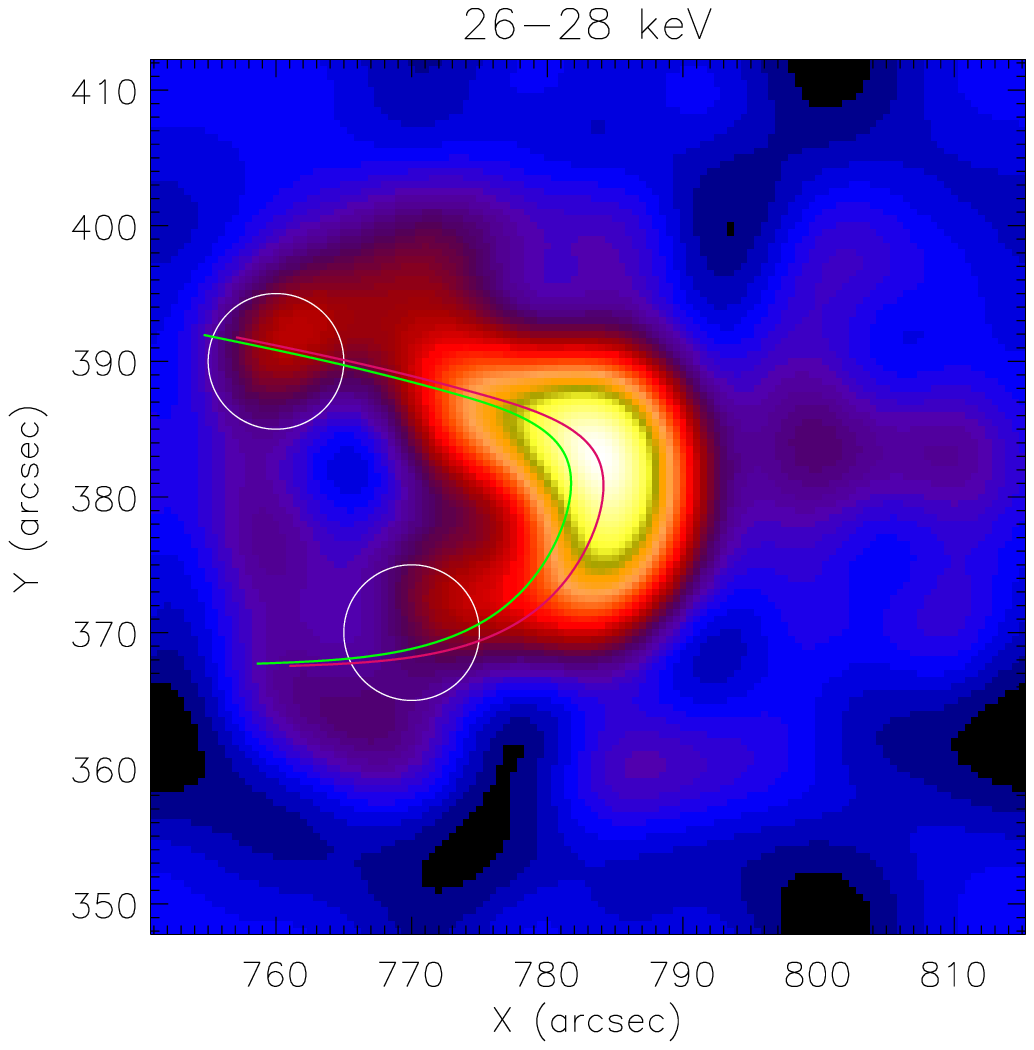}
  \includegraphics[width=0.24\textwidth]{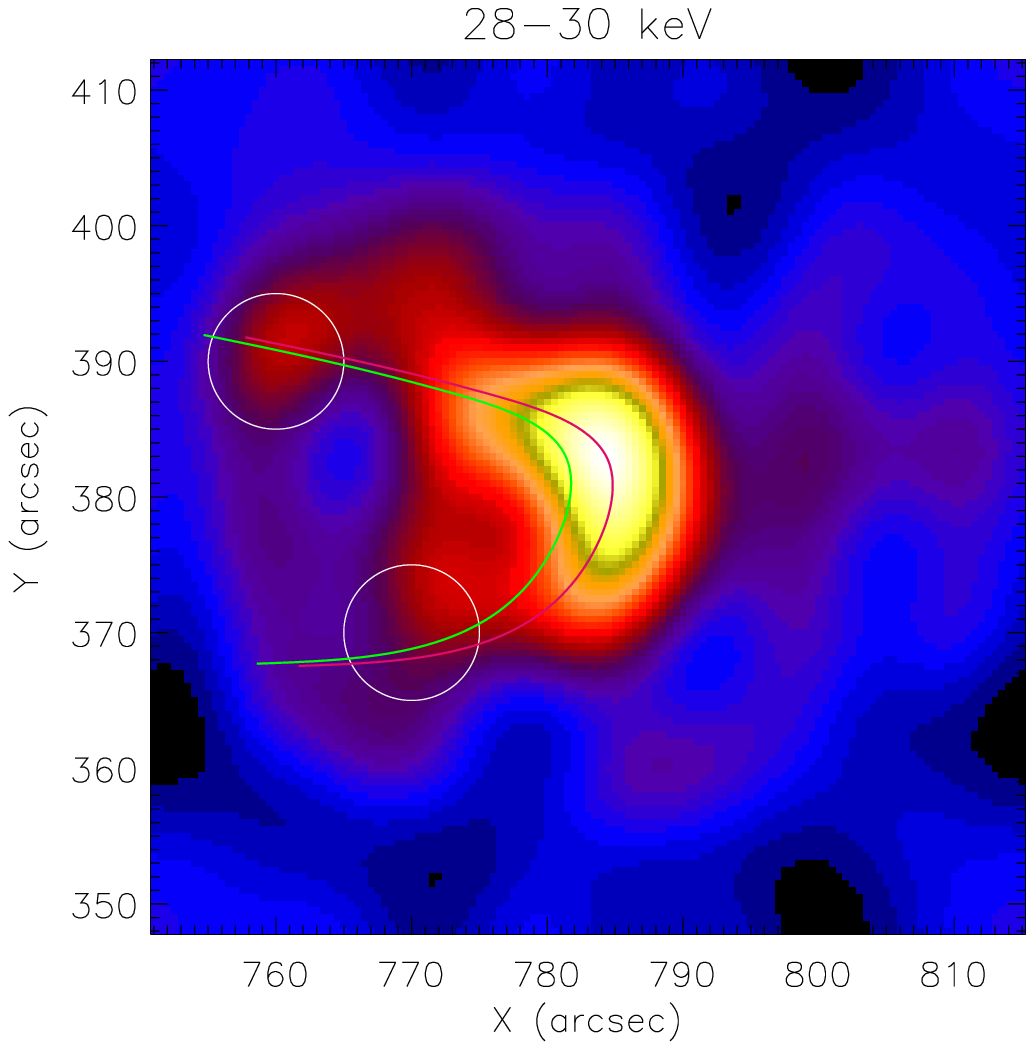}
\caption{RHESSI photon images for the 5-minute time interval on 15 April 2002 from 00:00 to 00:05 UT as shown in Figure \ref{Fig-lc15April2002}. The ten images are for 2 keV wide energy bins between 10 and 30 keV. They were made using normalized and regularized photon visibilities for detectors \#3, 4, 5, and 6 and the MEM NJIT image reconstruction algorithm with the ``tolerance'' parameter set to the default value of 0.03.  The two small circles in each image show the presumed location of two footpoints, and the green and purple arcs indicate the ``spine'' of the loop, as detailed in the text. }
\label{Fig-im15April2002}
\end{figure}

Photon images are shown in Figure \ref{Fig-im15April2002} for multiple 2-keV wide bins from 10 to 30 keV. They were made using the MEM\_NJIT reconstruction technique \citep{2007SoPh..240..241S}, with a tolerance of 0.03. Further, to produce images that vary smoothly from one energy channel to the next, and so reduce the scatter of the estimated loop length from one energy bin to the next, we used {\it regularized photon visibilities}. These are constructed by first finding the regularized electron visibilities \citep{2007ApJ...665..846P} corresponding to the count visibilities determined from count rates in the front segments of all detectors (except \#1 and \#2), and then forward-processing these electron visibilities to obtain more smoothly-varying photon visibilities. The visibilities were normalized to take into account the small differences in sensitivities of the different detectors for each event compared to the default values.

The images in Figure \ref{Fig-im15April2002} show the general loop-like appearance at all energies between 10 and 30 keV.  There is evidence for footpoint emission seen in Figure \ref{Fig-im15April2002-12to50kev} at energies above $\sim$20~keV, particularly from the northern footpoint.  Also evident are the above-the-loop-top sources reported by \citet{2004ApJ...612..546S} in the images between 10 and 22~keV.

Unfortunately, there are no EUV images from TRACE at this time, and so we cannot confirm that these apparent footpoint sources are actually on flare ribbons. Nevertheless, we have proceeded on the assumption that they are chromospheric footpoint sources and have developed techniques to ensure that they do not compromise the measurements of the length of the coronal part of the loop.  Clearly, this footpoint emission would significantly compromise the results presented by \citet{Guo2012,Guo2012a,Guo2013}, since they determined the source length by taking the second moment of all the emission in the field of view.

\begin{figure}
	\centering
    \includegraphics*[width=0.414\textwidth]{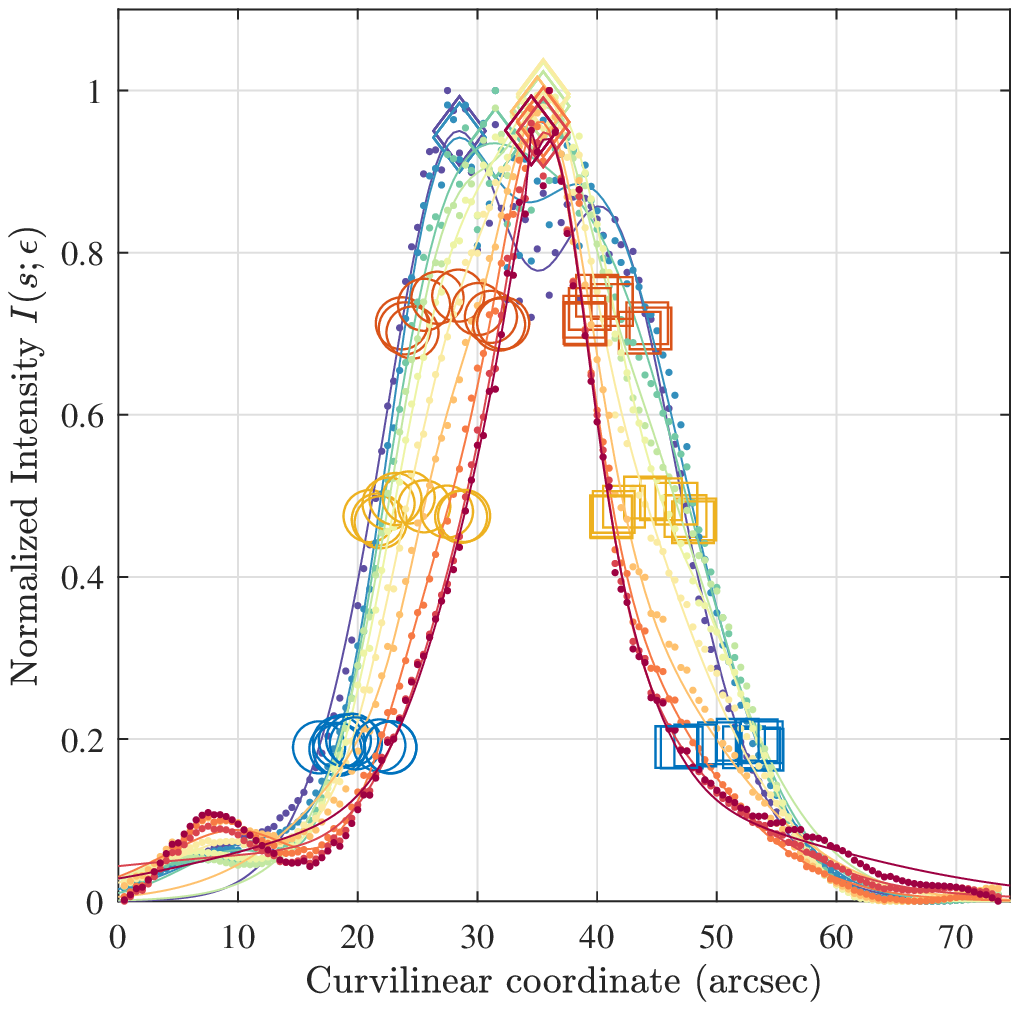}\
	\includegraphics*[width=0.536\textwidth]{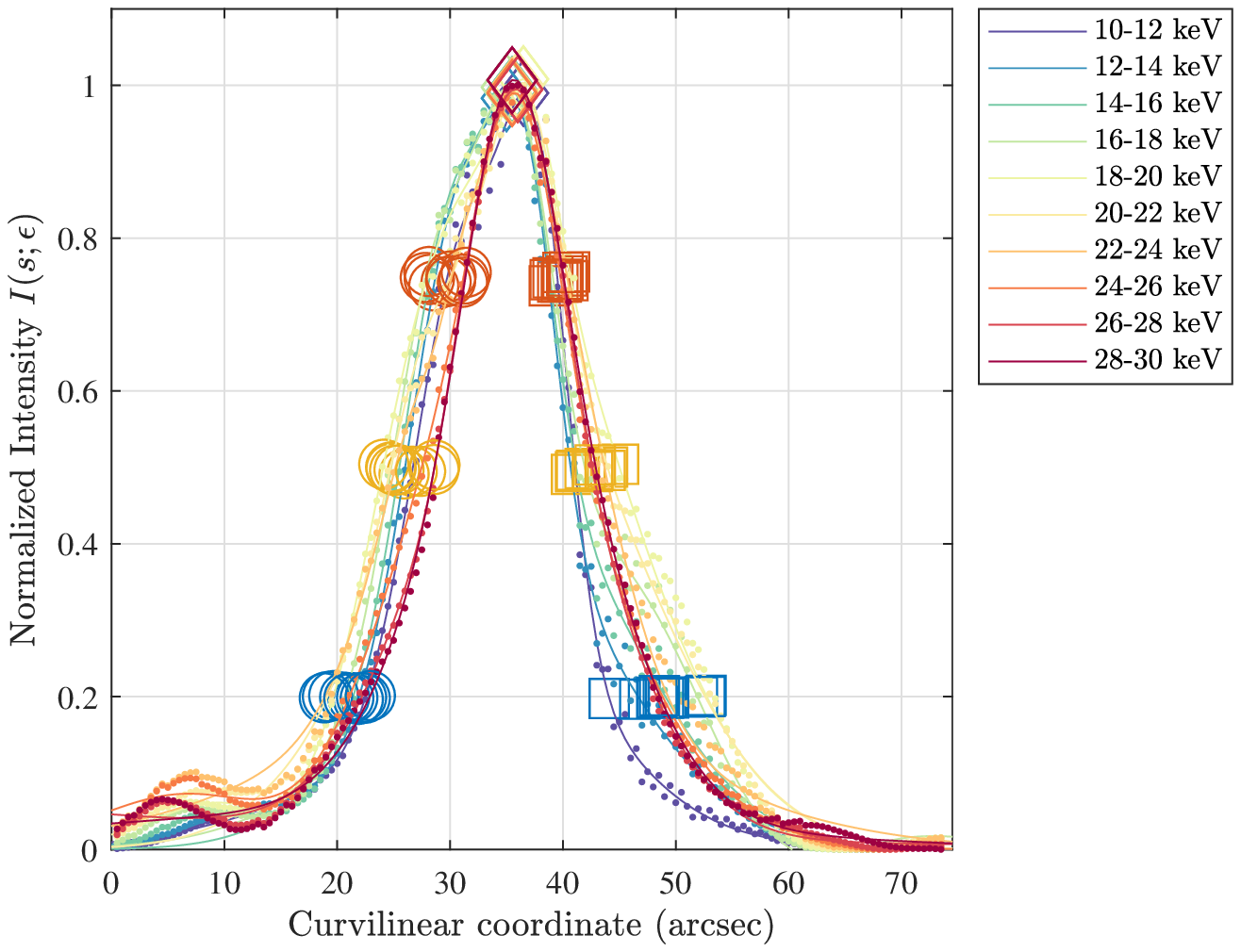}
\caption{\textbf{Left:} Normalized intensity vs.~distance measured from east to west along the fixed green arcs in Figure~\ref{Fig-im15April2002}. \textbf{Right:} Similar intensity vs.~distance for the energy-dependent purple arcs in Figure~\ref{Fig-im15April2002} that pass through the coronal source peak location at each energy.  Color-coded points for each energy show the intensities at each pixel along the arc in the image. The curves are the sum of the minimum number of Gaussians needed to adequately fit the data points. For each energy range, the diamonds show the location of the peak intensity; the circles and squares show the left and right location, respectively, at a given fraction of the peak intensity, with blue at 20\%, orange at 50\%, and red at 75\%.}
\label{Fig-IvsCC}
\end{figure}

For each flare time interval, we proceeded as follows to determine the geometric length of the coronal source independent of the footpoint emission:

\begin{itemize}

\item First, we made a 20--30~keV image for the first time interval of each flare to determine the footpoint locations to within $\sim$5~arcsec in X and Y. These are given in Table~\ref{tab-parameters}. We used the same footpoint locations for subsequent time intervals of the same flare. For flares where footpoints could not be reliably located independently of the coronal source, dashes are shown in Table \ref{tab-parameters} and no further analysis was carried out.

\item Using both the shape of the coronal source in the 20--22~keV image and the location of the footpoints, we constructed a locus of points passing through pixels with the brightest emission along the ``spine'' of the coronal source and extending through the footpoints. This spine is shown as the green arc in each image of Figure \ref{Fig-im15April2002} at the same location for all energies. 

\item Next, for each energy bin we moved the green arc, without changing its shape, so that it passes through the point of peak emission in that image. The distance and direction of the move is the difference between the location of the peak emission in the 20--22~keV image and the location of the peak in that energy bin. We interpret these new arcs, shown in purple in each image, as delineating the magnetic field line about which the electrons spiral. Note that the purple arc, in general, still passes close to the two footpoints. 

We found that the location of the purple arcs with respect to the green arcs was energy dependent, with the purple arcs situated to the left of the green arcs at lower energies and to the right at higher energies. Given that this flare occurred near the western limb of the Sun ($X$ = 781~arcsec), this corresponds to an increase in altitude of the coronal source with increasing energy.

\item For both the green and purple arcs in each energy bin ($\epsilon$), we measured the photon intensity $I(s;\epsilon)$ of the extended coronal source along the line in question using the curvilinear coordinate $s$ measured from the north-east end of the arc. The curves of $I(s;\epsilon)$ vs.~$s$ are shown in Figure~\ref{Fig-IvsCC} for each energy bin. The coronal source is seen as the major peak centered at $s \simeq$~35~arcsec with the two much weaker footpoints seen at the higher energies centered at $s \simeq$~7 and $\simeq$~62~arcsec.

\end{itemize}

\subsection{Determination of source length}
	\label{sub-length}

As mentioned in Section~\ref{sec:intro}, previous authors \citep{2008ApJ...673..576X,Guo2012,Guo2012a,Guo2013} used an integral moment calculation to estimate the source length $L$ in each photon energy bin. However, such methods give substantial weight to any footpoint emission that may be present. Since the spectrum of the footpoint emission is generally significantly harder than that of the coronal source, the source length calculated in this way can increase with energy and will not reflect the true variation of the coronal source length.  To be sure that we did not include any footpoint emission in our estimate of the coronal source length, we adopted a different methodology, as follows:

\begin{enumerate}

\item Ensure that the green and purple arcs defined above and shown in each image of Figure \ref{Fig-im15April2002} passed through, or within a few arcsec of, the presumed locations of two footpoints seen at higher energies.


\item Generate the plots of intensity \textit{vs.}~distance along the green and purple arcs shown in Figure \ref{Fig-IvsCC}.

\item At each photon energy $\epsilon$, compute the sum of a minimum number of Gaussians to adequately fit the data (i.e.,~to give a reduced $\chi^2 < 1.5$). This allows for possible asymmetry (skewness) in the form of $I(s;\epsilon)$ for the coronal source and also emission from the footpoints.

\item Locate the dominant peak in $I(s)$ closest to the center of the arc at each energy and determine the points along the arc where the intensity decreased to 75\%, 50\%, and 20\% of the peak intensity.  

\item For each energy bin, determine the distances along the arc between the two points at which the normalized intensity dropped to each of the three percentages of the peak. 

\end{enumerate}

\subsection{Results}
	\label{sub-results}
    
\begin{figure}
	\centering
    \includegraphics*[width=0.45\textwidth]{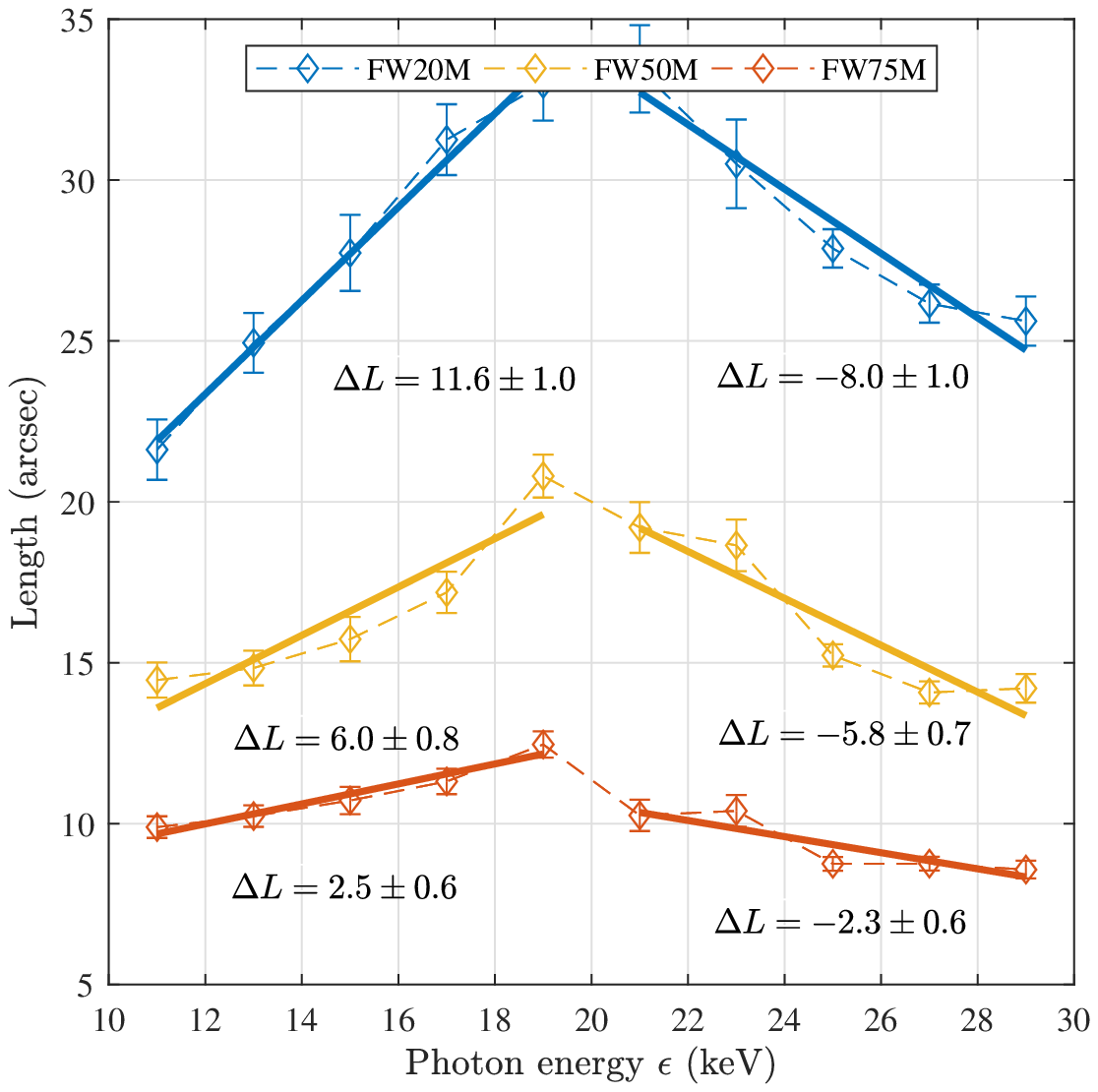}
	\includegraphics*[width=0.45\textwidth]{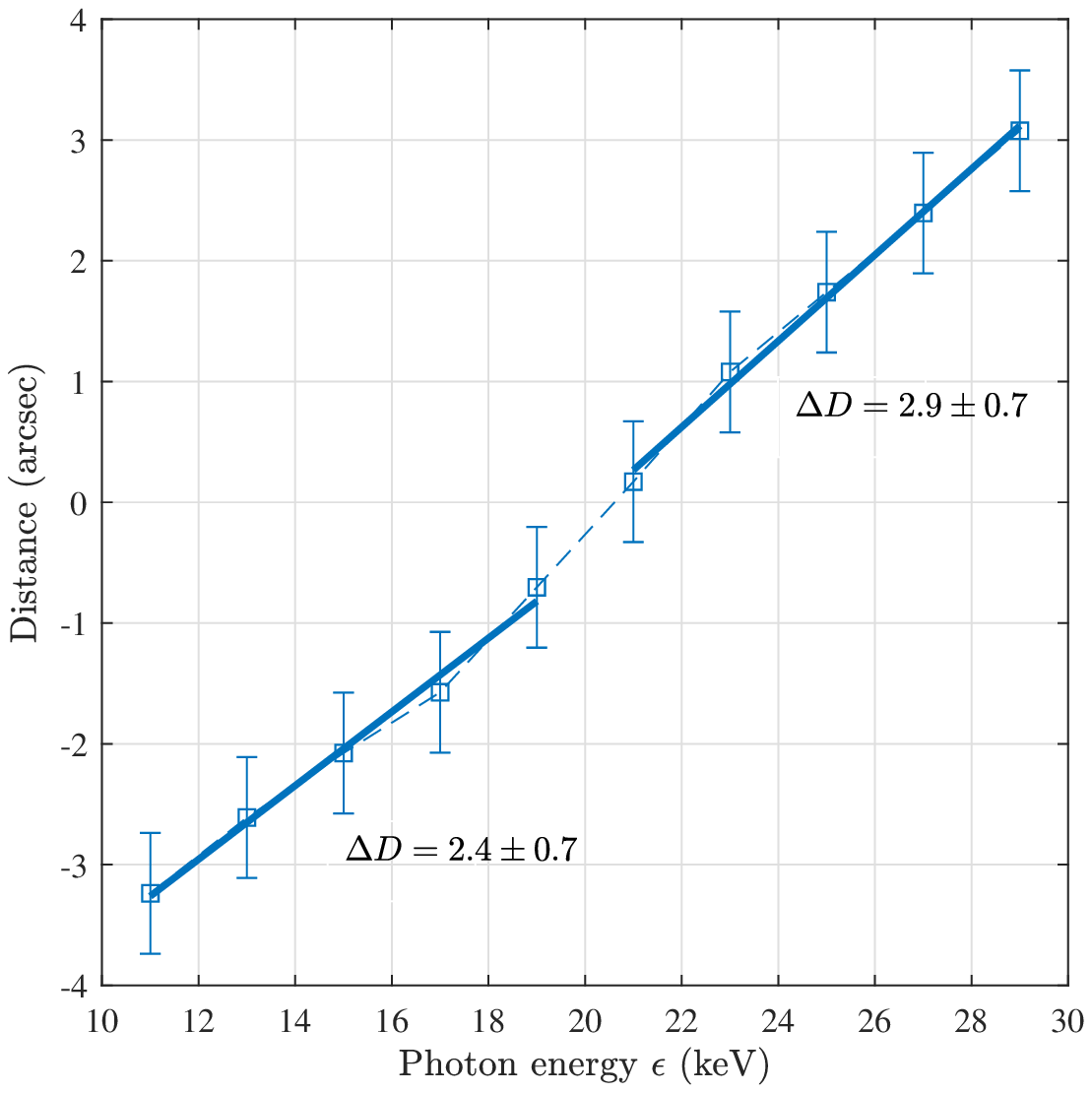}
\caption{\textbf{Left:} Coronal source length measured along the energy-dependent purple arcs in Figure~\ref{Fig-im15April2002} plotted vs.~photon energy. 
The lengths were determined as the distances between the positions at which the intensity along the arc dropped to 75\% (blue), 50\% (orange), and 20\% (red) of the peak value in each energy bin. Error bars show the $\pm1~\sigma$ uncertainties derived from the Gaussian fits shown in Figure \ref{Fig-IvsCC} but limited by the image pixel size used to a minimum of 0.5 arcsec.. 
The Solid lines indicate the best fit to the data points from 10 to 20 keV and from 20 to 30 keV. The changes in length ($\Delta  L$) over these two energy ranges are shown on the plot, and listed in Table \ref{tab-parameters}. 
\textbf{Right:} Energy dependence of the distance, $D$, between the peak in flux along the green arc in the 20--22~keV image and the peak along the purple energy-dependent arcs in Figure~\ref{Fig-im15April2002}. 
The solid lines show linear fits to the data points from 10 -- 20~keV and 20 -- 30~keV, and the indicated values of $\Delta D$ (listed in Table~\ref{tab-parameters}) are the changes in position in arcsec over the two energy ranges.}
\label{Fig-LvsE_HvsE}
\end{figure}

Figure \ref{Fig-LvsE_HvsE} (Left) shows the coronal source length as a function of energy determined from the images along the energy-dependent purple arcs shown in Figure~\ref{Fig-im15April2002}. Three different estimates of the source length are plotted as determined from taking the width of the distribution along the arcs at the three different levels below the peak intensity - 75, 50, and 20\%.  In no case is there evidence for the increase in loop length with increasing energy above $\sim$20~keV reported by \citet{Guo2012,Guo2012a,Guo2013}. 
To parameterize the change in the source length with photon energy, L($\epsilon$), we have made linear fits between 10 and 20~keV, and between 20 and 30~keV, as shown in Figure \ref{Fig-LvsE_HvsE} (Left). The changes in length $\Delta$L over these two energy ranges are indicated in the plot for the three different fractions of the peak intensities.  In each case, there is evidence of an increase in loop length for the thermal source but a decrease in length for the nonthermal source at energies above 20 keV.  This is in contrast to the \textbf{increase} in loop length of as high as 10 arcsec~between 15 and 25 keV reported by \citet{Guo2012} for this same time interval (estimated from Figure 4 of that paper).

In deciding what fraction of the peak value to use in order to best characterize the source length, we considered the need to provide a sufficient range of points to adequately determine the width of the $I(s)$ profiles in Figure \ref{Fig-IvsCC}. The intensities should be well above the noise in the image from the statistics and from the limitations of the image reconstruction. Consistent results were found for most events for each of the three percentage levels that we used but for some events with stronger footpoints at the higher energies (particularly the X1.2 flare on 15 May 2013), the 20\% level made it more difficult to separate the coronal and footpoint emission along the arcs. The 75\% level gives a measure of only the top of the coronal source. Finally, we have chosen to use the 50\% level in reporting the changes $\Delta L$ for all events in Table \ref{Tab-AltSummary}. We believe that this gives the best estimate of the coronal source length unaffected by any footpoint emission. It also has the advantage of being directly comparable with the 1$\sigma$ and FWHM source lengths reported by \cite{2008ApJ...673..576X}, \cite{Guo2012}, and \cite{2011ApJ...730L..22K}.

\begin{figure}
	\centering
    \includegraphics[width=0.7\textwidth]
	{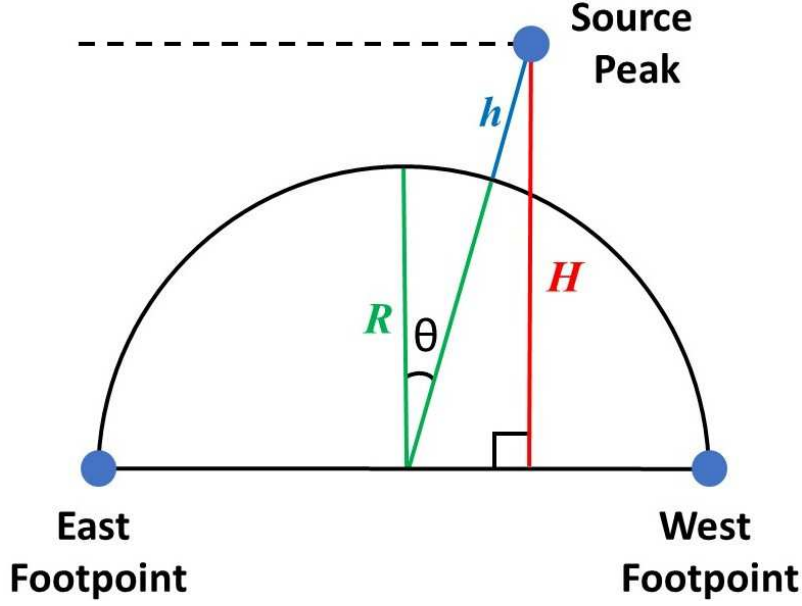}\
\caption{Diagram showing the definition of the source height, $H$, above the photosphere and the height, $h$, above a vertical semicircular loop with radius, $R$, between two footpoints separated by $2R$; $\theta$ is the angle between the vertical and the line from the source peak location to the center of the semicircular loop through the two footpoints.}
\label{Fig-AltitudeDiagram}
\end{figure}

In addition to the source lengths at different energies obtained from the images, we also determined the positions of the peak emission along the purple arcs shown in Figure~\ref{Fig-im15April2002}. These are shown as a function of photon energy in Figure \ref{Fig-LvsE_HvsE} (Right). 
Note that points at adjacent energies are not statistically independent because of the smoothing of the emitting electron spectrum imposed by the regularization technique used to construct the images \citep{2007ApJ...665..846P}. The change in position with energy is clearly seen over the full energy range with a change $\Delta D$ of 2.4~arcsec between 10 and 20~keV and 2.9~arcsec~between 20 and 30~keV.  Values of the change in position over these energy ranges are given in Table \ref{Tab-AltSummary} for each time interval analyzed. If these changes of position are interpreted as a change in the altitude of the source, then $\Delta H$ is $(1.3\pm0.1)$~Mm between 10 and 20~keV and $(1.9\pm0.1)$~Mm between 20 and 30~keV. For the conversion from a position on the solar disk to a source altitude, we assumed that the coronal source was located vertically above the footpoints (see Figure  \ref{Fig-AltitudeDiagram}, and the correction was applied for the foreshortening resulting from the flare location on the solar disk. 



We tried two other methods for determining the length of the coronal source in each image. The first was to make images in each of the same 2-keV energy bins using the forward fitting image reconstruction method \citep{2002SoPh..210...61H} called VIS\_FWDFIT currently available in the RHESSI IDL software in Solar Soft (SSW).  We used two circular footpoint sources with fixed locations but variable intensities plus a single loop with variable location, intensity, length, width, and curvature. In this method the free source parameters are adjusted until a minimum is achieved in chi-squares determined from a comparison between the measured and calculated visibilities. In this way, we were able to obtain the coronal source parameters independently of any footpoint emission. The second method was adapted from the scheme described by \cite{1999ApJ...515..842A} in which an assumed semicircular loop is projected onto the plane of the sky at the location of the flare on the solar disk.\footnote{The basic IDL code is available in Solar Software at \textbackslash ssw\textbackslash packages\textbackslash hydro\textbackslash idl with a tutorial at 
 \url{http://www.lmsal.com/~aschwand/hydro_software/hydro_tutorial1.html}.} The loop dimensions, orientation, and intensity are allowed to vary to give a least-squares fit to the MEM\_NJIT image in each energy range, excluding emission from the presumed footpoint locations. Both of these methods gave results for the 14/15 April 2002 event (\#3, 4, and 5 in Table~\ref{tab-parameters}) similar to those obtained by the method described in Section \ref{sub-goemetry} and \ref{sub-length}, so that method was adopted for all events analyzed.
 

\section{Summary of Coronal HXR source parameters}\label{summary-of-parameters}

\begin{figure}
	\centering
    \includegraphics*[height=0.4\textwidth]{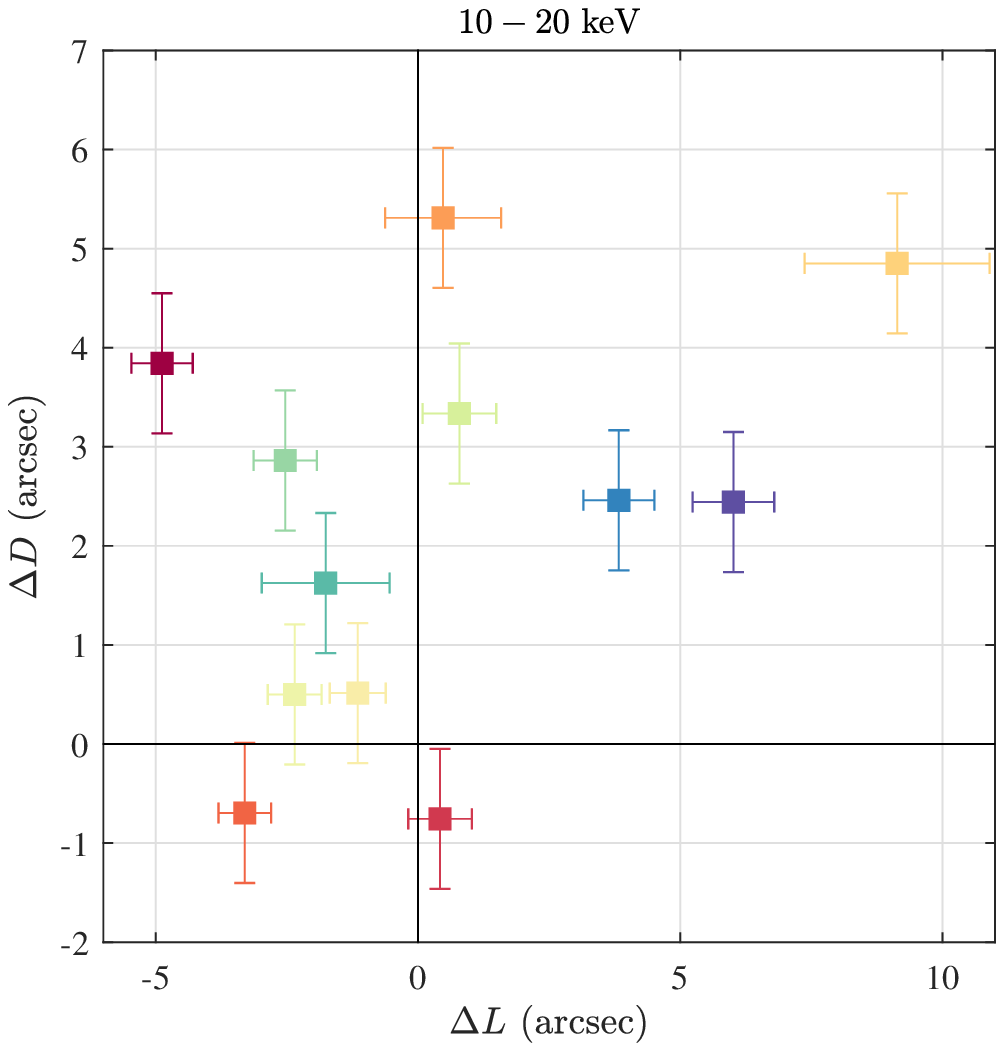}
    \includegraphics*[height=0.4\textwidth]{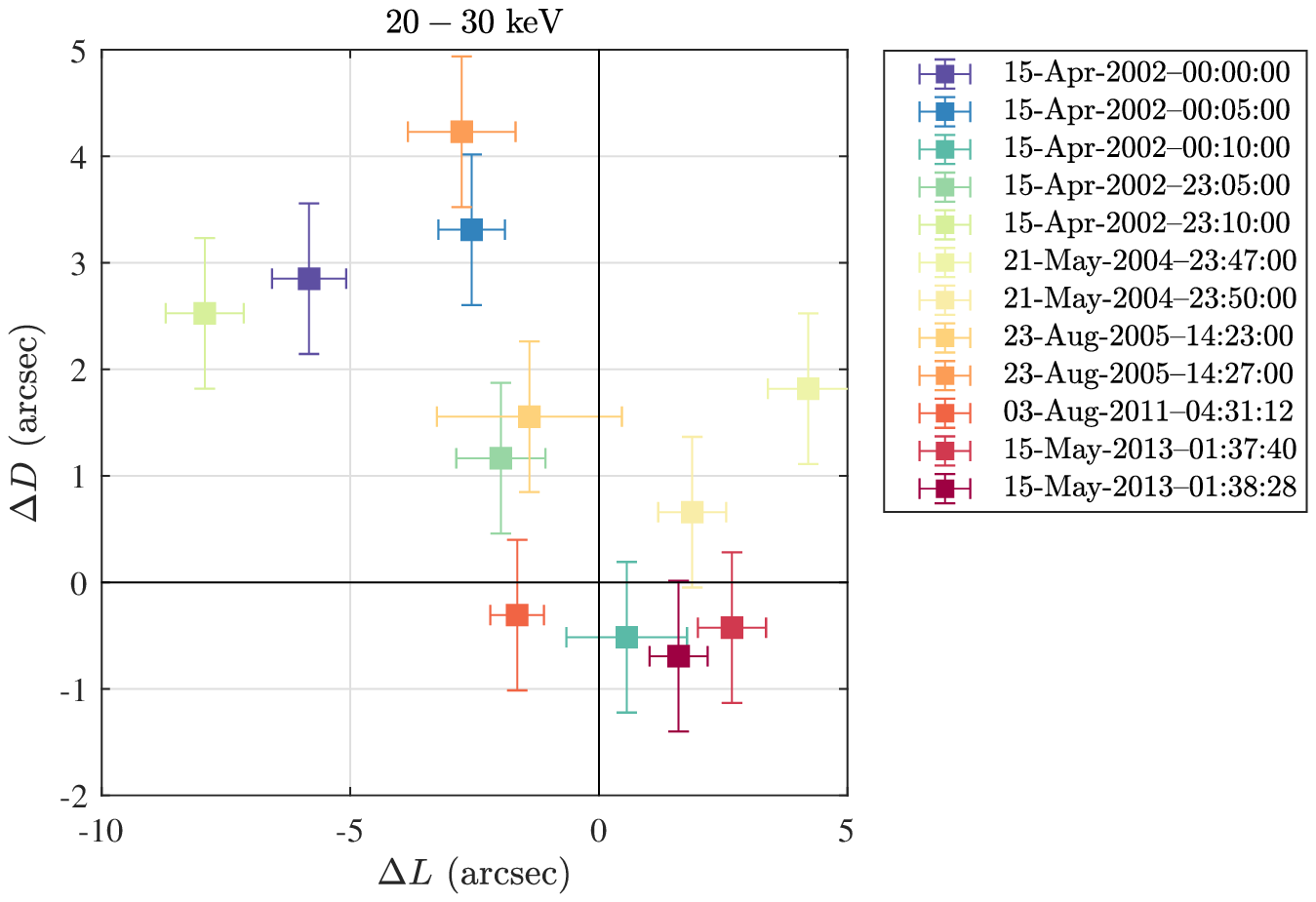}
\caption{Estimated changes in position ($\Delta D$) and length ($\Delta L$ measured at 50\% of the peak intensity) between 10 and 20 and 20 to 30 keV for the coronal sources in all the time intervals used in the study listed in Table \ref{Tab-AltSummary}.
}
\label{Fig-DHvsDL}
\end{figure}

In this section, we present the results for just those events and time intervals for which we determined that the coronal source could be reliably separated from the footpoint emission, i.e.~those listed with a ``Y'' in the last column of Table~\ref{tab-parameters}. \textbf{Values determined for the derived parameters for each of these time intervals are listed in Table \ref{Tab-AltSummary}, with the mean value and the standard deviation of the scatter of the values in the last two rows of each column. A plot of the change in source position ($\Delta D$) vs.~the change in loop length ($\Delta L$) between 10 and 20 keV, and between 20 and 30~keV, is shown in Figure~\ref{Fig-DHvsDL}.} 

\textbf{For the 6 selected events and a total of 12 time intervals, the mean change in source length ($\Delta L$) with energy between 10 and 20 (20 and 30) keV is $-0.9 \pm 0.2$ (-1.0 $\pm$0.2) arcsec with a standard deviation of 4.1 (3.5) arcsec. Increases in $\Delta L$ are always less than $\sim$3~arcsec except for three cases in the 10 - 20~keV energy range (events \#3 and 4 on 15~April~2002 on \#21 on 23 August 2005) and just one event in the 20 - 30 keV range (\#14 on 21 May 2004). Decreases in source length with energy of $>$3 arcsec are measured in 2 cases in the 10 - 20 keV range (\#25 on 03~August 2011 and \#28 on 15~May 2013) and in 2 cases in the 20 - 30 keV range (\#3 on 15 April 2002 and \#7 on 15 April 2002). The mean change in position $\Delta D$ with energy between 10 and 20 (20 and 30) keV is 2.2 (1.4) arcsec with a standard deviation of 2.0 (1.6) arcsec. The values of $\Delta D$ are all consistent with zero or a positive value as high as 5.3~arcsec in both the 10 - 20 keV and 20 - 30 keV energy ranges, corresponding to no change or to an increase in altitude with energy. Note that in all cases, the coronal source is significantly above a postulated vertical semicircular loop, i.e.~$h$ is positive by as much as 27 arcsec corresponding to 20 Mm in the case of event \#21 on 23~August~2005 and events 27 and 28 on 15~May~2013.}

\begin{table}
 \centering
 \begin{tabular}{|r|c|c|c|c|c|c|c|c|c|}
  \hline
  \hline
    \multirow{3}{*}{\#} & \multirow{3}{*}{Date} &\multicolumn{2}{c|}{$\Delta L$ arcsec} & \multicolumn{2}{c|}{$\Delta D$ arcsec} & {$R$}  &{$H$}  & {$h$}           & \multirow{3}{*}{$\theta$ ($^\circ$)}\\
  \cline{3-9}
\   &   & 10-20 & 20-30 & 10-20 & 20-30 & \multicolumn{3}{c|}{\multirow{2}{*}{arcsec/Mm}} & \\       
  \   &   & keV   & keV   & keV   & keV   & \multicolumn{3}{c|}{~}         & ~         \\
  \hline
    3 & 15-Apr-2002 & 6.0$\pm$0.8 & -5.8$\pm$0.7 & 2.4 & 2.9 & 11/8 & 18/13 & 7/6 & 15 \\
    \hline 
    4 & 15-Apr-2002 & 3.8$\pm$0.7 & -2.6$\pm$0.7 & 2.5 & 3.3 & 11/8 & 20/15 & 9/7 & 13 \\
    \hline
    5 & 15-Apr-2002 & -1.8$\pm$1.2 & 0.6$\pm$1.2 & 1.6 & -0.5 & 11/8 & 23/17 & 13/10 & 20 \\
    \hline
    6 & 15-Apr-2002 & -2.5$\pm$0.6 & -2.0$\pm$0.9 & 2.9 & 1.2 & 22/17 & 18/14 & 2/2 & 43 \\
 \hline
    7 & 15-Apr-2002 & 0.8$\pm$0.7 & -7.9$\pm$0.8 & 3.3 & 2.5 & 22/17 & 16/12 &  5/4 & 54 \\
    \hline
    14 & 21-May-2004 & -2.4$\pm$0.5 & 4.2$\pm$0.8 & 0.5 & 1.8 & 12/9 & 18/14 &  7/5 & 10 \\
    \hline
    15 & 21-May-2004 & -1.2$\pm$0.5 & 1.9$\pm$0.7 & 0.5 & 0.7 & 12/9 & 17/14 &  5/5 & 9 \\
    \hline
    21 & 23-Aug-2005 & 9.1$\pm$1.8 & -1.4$\pm$1.9 & 4.9 & 1.6 & 21/16 & 47/35 & 27/20 & 12 \\
    \hline
    22 & 23-Aug-2005 & 0.5$\pm$1.1 & -2.8$\pm$1.1 & 5.3 & 4.2 & 21/16 & 43/32 & 22/17 & 7 \\
    \hline
    25 & 03-Aug-2011 & -3.3$\pm$0.5 & -1.6$\pm$0.5 & -0.7 & -0.3 & 8/6 & -- & -- & -- \\
    \hline
    27 & 15-May-2013 & 0.4$\pm$0.6 & 2.7$\pm$0.7 & -0.8 & -0.4 & 11/8 & 38/28 & 27/19 & 3 \\
    \hline
    28 & 15-May-2013 & -4.9$\pm$0.6 & 1.6$\pm$0.6 & 3.8 & -0.7 & 11/8 & 38/29 & 27/20 & 2 \\
   
  \hline
  \hline
\multicolumn{2}{|r|}{\textbf{Mean}}       & -0.9$\pm$0.2 & -1.0$\pm$0.2 & 2.2 & 1.4 & 14/11 & 27/20 & 14/10 & 17 \\
 \hline
 \multicolumn{2}{|r|}{\textbf{Standard Deviation}} & 4.1          & 3.5          & 2.0 & 1.6 &  5/4  & 12/9 & 10/7 & 17 \\
  \hline
  \hline
   \end{tabular}
 \caption{Results of the analysis for the event time intervals in Table~\ref{tab-parameters} for which the footpoint locations can be identified and the coronal source can be reliably separated from any footpoint emissions. The columns are as follows: \# is the interval number from Table~\ref{tab-parameters}, Date is the date of the time interval, $\Delta L$ is the change in the FWHM length of the coronal source between 10 and 20~keV and 20 to 30~keV, $\Delta D$ is the change in position of the coronal source peak in the same energy ranges with 1$\sigma$ uncertainties of $\pm$0.7 arcsec, $R$ is the radius of a semicircular loop drawn through the identified footpoints listed in Table~\ref{Tab-AltSummary}, $H$ is the height of the 20 keV coronal source above the photosphere, \textbf{$h$ is the distance from the coronal source to the semicircular loop}, and $\theta$ is the angle to the local vertical (see Figure~\ref{Fig-AltitudeDiagram}). The values of $R$, $H$, $h$, and $\theta$ are for the source peak location in the 20 - 22 keV image and are corrected for the foreshortening from the location of the event on the solar disk. All distances are in arcsec with the values of $R$, $H$, and $h$ also given in Mm. Uncertainties on $R$, $H$, and $h$ are typically $\pm$5 arcsec determined from the accuracy with which the footpoint and coronal sources can be located. Note that the event on 03 August 2011 was too close to disc center to allow reliable source height determinations to be made.}
 \label{Tab-AltSummary}
\end{table}

\section{Summary and Conclusions}\label{conclusions}

Recognizing the possible influence of (even faint) footpoint emission on the length of a source calculated using integral moments, we have carried out a re-analysis of the events studied by \citet{2004ApJ...603L.117V}, \cite{2008ApJ...673..576X}, and \citet{Guo2012,Guo2012a,Guo2013}, together with a more recent event.  In our approach, we have minimized the possible influence of footpoint emission in three ways: (1) by defining a loop ``spine'' passing through the presumed locations of the two footpoints and through the peak of the coronal source, (2) by using context information from EUV images, when available, to better determine the possible locations of footpoints, and (3) by defining the ``length'' of a coronal source in a manner that uses the ``shape'' of the intensity-versus-position curve, rather than integral moments, which are strongly influenced by the tails of the $I(s;\epsilon)$ profiles at relatively large distances from the intensity peak. Specifically, we use a strict intensity cutoff to estimate the ``shape parameter'' of a fitted Gaussian form, i.e., the full width at 50\% of the peak.

Using this revised methodology, we find that the previously-inferred increase in source length with energy is no longer apparent. Further, we also find that coronal sources at higher photon energies generally appear at higher altitudes, indicating that hot plasma at different temperatures and accelerated electrons at different energies must be on different field lines. Thus, the original postulate of \citet{2008ApJ...673..576X} that the coronal HXR sources seen at different energies are all in the same magnetic loop is not tenable.



Although we have found that $L(\epsilon)$ does not increase, neither does it {\it decrease}, as would be expected for a compact thermal source, with higher energies corresponding to more central regions in the source \citep{2008ApJ...673..576X}.  We therefore believe that the inferred source length $L$ represents, to a large degree, the combined extent of the various energy release and acceleration regions, and that the HXR emission represents a combination of thermal and nonthermal components.  The density in the acceleration region can be estimated from the soft X-ray emission measure and/or from comparison of the spectra in the coronal source and at the footpoints (\citealp[cf.][]{2013A&A...551A.135S}).  Thus, analysis of such coronal HXR sources can still provide us with the information necessary to calculate the number ${\cal N} = nAL$ of electrons available for acceleration and hence the specific acceleration rate \citep{2008AIPC.1039....3E,Guo2012a}, a very useful measure of the efficiency of the electron acceleration process that can be compared with candidate acceleration models. However, to determine realistic densities, this analysis ideally requires a knowledge of the differential emission measure to lower temperatures than covered by RHESSI \citep{2015A&A...584A..89J}.  This has been attempted by \citet{2013ApJ...779..107B,2014ApJ...789..116I,2015ApJ...815...73B,2018ApJ...856L..17S} using data from AIA and by \citet{2014ApJ...788L..31C} for larger events using data from the EUV Variability Experiment (\citealp[EVE,][]{2012SoPh..275..115W}) on SDO to cover the lower temperatures.  However, AIA and EVE data are only available after SDO was launched in 2010 and such a detailed analysis is beyond the scope of this paper.   

At least for the thermal emission, the apparent increase in altitude with increasing energy is consistent with the standard CSHKP flare model \citep{1964NASSP..50..451C,1966Natur.211..695S,1974SoPh...34..323H,1976SoPh...50...85K} \textbf{reviewed by \cite{2002A&ARv..10..313P} and} recently updated by 
\citet{2016JGRA..12111667H}. In this model, reconnection takes place in a current sheet above the previously generated cooling flare loops.  As the flare progresses in time, and reconnection continues, new hot loops are formed at higher altitudes.  The previously formed loops at lower altitudes will have cooled to a lower temperature than that of the newly formed loops and hence will have a softer spectrum. Thus, the peak or centroid of the ensemble of loops at any given time will be at a higher altitude for higher energies. This purely ``thermal'' explanation of new hot loops forming at higher altitudes was used by \cite{2003ApJ...596L.251S} but they used emission up to only a 16–-20 keV energy bin that was almost certainly thermal. Similarly, \cite{2004ApJ...612..546S} used the same ``thermal'' explanation for the increase in source altitude with energy during three April 2002 flares but again it is likely that the emission in the energy ranges (6--12 and 12--25 keV) that they imaged was predominantly thermal. They do show two images in the 25--50 keV energy range that also show the coronal source at a higher altitude than the lower energy source suggesting that the increase in altitude extends into the nonthermal domain.

For the nonthermal component, the explanation is not so clear. The separation between the thermal component at low energies and the nonthermal component at higher energies is not uniquely defined since it depends on the form of the differential emission measure at high temperatures.  For the spectral fit in Figure~\ref{Fig-sp15April2002}, where the assumed DEM was a power law in temperature, the two components have equal intensity at 22 keV and there is a significant thermal component up to the 30 keV maximum energy that we could use for decent images.  Thus, it is surprising that the increase in source peak altitude shown in the right panel of Figure~\ref{Fig-LvsE_HvsE} continues to increase with energy at the same rate over the full energy range covered. This question needs further study but examination of the values of $\Delta D$ in Table \ref{Tab-AltSummary} shows that this is not always the case.

Finally, we note that most of the events show evidence for both thick--target coronal sources and chromospheric footpoints visible at higher energies. Thus, the basic collisional transport model of \citet{2008ApJ...673..576X,Guo2012,Guo2012a,Guo2013} and \citet{2015A&A...584A..89J} must be modified to include escape of high--energy electrons from the coronal source in order to create chromospheric footpoints. Just such an extension was considered by \citet{2014ApJ...796..142B}. Comparison of this enhanced model with RHESSI data, complemented as appropriate with data from SDO, STEREO, and Hinode, will be the focus of future work.


\acknowledgements
\begin{center}ACKNOWLEDGEMENTS\\\end{center}

All RHESSI data analysis was carried out using both the graphical user interface (GUI) and the command-line interface to the IDL routines in Solar Software (SSW). The image reconstructions were made using the techniques discussed by  \citet{2002SoPh..210...61H} and \citet{2007SoPh..240..241S}. The spectroscopy was done with the object-oriented OSPEX package in SSW discussed by \citet{2002SoPh..210..165S}. Albert Shih (GSFC) is acknowledged for his help in making the corrections for foreshortening. AGE was supported by grant NNX17AI16G from NASA's Heliophysics Supporting Research program.

\appendix

\section{Comments on Analyzed Events}
	\label{Appendix}

\begin{description}

\item [12 April 2002 17:27 to 18:13 UT] ~\\ Footpoint locations are indeterminate so loop and footpoint sources cannot be reliably separated. Studied by \citet{2008ApJ...673..576X,Guo2012,Guo2012a,Guo2013,2018arXiv180309847F}.

\item [14 April 2002 23:54 UT to 15 April 2002 00:41 UT] ~\\ Both footpoints identified and separable from the loop source source. Studied by \citet{2008ApJ...673..576X,Guo2012,Guo2012a,Guo2013}.

\item[15 April 2002 22:54 to 23:21 UT] ~\\Both footpoints identified and separable from the loop source. Studied by \citet{2004ApJ...603L.117V,2008ApJ...673..576X,Guo2012,Guo2012a,Guo2013}.
Above the looptop sources identified by \citet{2003ApJ...596L.251S} and \citet{2004ApJ...612..546S}.

\item[17 April 2002 16:50 to 17:11 UT] ~\\
Footpoint locations are indeterminate so loop and footpoint sources cannot be reliably separated. Studied by \citet{2008ApJ...673..576X,Guo2012,Guo2012a,Guo2013}.

\item[17 June 2003 22:22 to 23:07 UT] ~\\
Three possible HXR footpoint sources. Coronal source not detectable above $\sim$24 keV. TRACE 195~\AA~images are available but they don't help to identify the footpoint and loop top sources. Studied by \citet{2008ApJ...673..576X,Guo2012,Guo2012a,Guo2013}.

\item[10 July 2003 14:13 to 14:31 UT] ~\\
RHESSI saw only the decay of this event. The footpoints were apparently occulted behind the west limb. TRACE 1600~\AA~images are available. Studied by \citet{2008ApJ...673..576X,Guo2012,Guo2012a,Guo2013}.

\item[02 December 2003 22:51 to 23:09 UT] ~\\
Limb event, footpoints occulted \textbf{so not included in our list of events.} Studied by \citet{2008ApJ...673..576X}.

\item[21 May 2004 22:32 UT to 22 May 2004, 00:21 UT] ~\\
Both footpoints identified and separable from the loop. Studied by \citet{2008ApJ...673..576X,Guo2012,Guo2012a,Guo2013}.
EIT 195~\AA~image is available.

\item[31 August 2004 05:20 to 05:44 UT] ~\\
Only detected to  $\sim$25 keV. Particle contamination evident in light curve. Studied by \citet{2008ApJ...673..576X,Guo2012,Guo2012a,Guo2013}.

\item[01 June 2005 02:33 to 03:12 UT] ~\\
HXR footpoints were $<$10 arcsec apart and not well separated from the coronal source. Studied by \citet{2008ApJ...673..576X,Guo2012,Guo2012a,Guo2013}.

\item[23 August 2005 14:05 to 14:49 UT] ~\\
Near limb event with both footpoints on the visible disc. Loop well separated. Studied by \citet{2008ApJ...673..576X}.

\item[13 February 2011 17:30 to 18:09 UT] ~\\
Footpoint locations could not be determined. Multiple loops appeared to be involved. Bright ribbons seen in AIA 1700~\AA~images. Studied by \citet{Guo2012,Guo2012a,Guo2013}.

\item[03 August 2011 04:29 to 04:44 UT] ~\\
Close to disk center and footpoints $<$10 arcsec apart so it is impossible to separate the footpoints from the loop. Studied by \citet{Guo2012,Guo2012a,Guo2013}.

\item[25 September 2011 03:25 to 03:42 UT] ~\\
Very narrow ($<$2 arcsec wide) line source on light bridge. Location of footpoints not clear. Possible multiple loops involved. AIA images are available. Studied by \citet{Guo2012,Guo2012a,Guo2013}.

\item[15 May 2013 01:37 to 01:43 UT] ~\\
X1 flare with two footpoint sources seen up to 50 - 100 keV and bright loops in AIA 131~\AA~images.

\end{description}




\bibliography{CoronalHXRSources,Mendeley07Dec2018}

\begin{thebibliography}{}
\expandafter\ifx\csname natexlab\endcsname\relax\def\natexlab#1{#1}\fi
\providecommand{\url}[1]{\href{#1}{#1}}

\bibitem[{{Aschwanden} {et~al.}(1999){Aschwanden}, {Newmark},
  {Delaboudini{\`e}re}, {Neupert}, {Klimchuk}, {Gary}, {Portier-Fozzani}, \&
  {Zucker}}]{1999ApJ...515..842A}
{Aschwanden}, M.~J., {Newmark}, J.~S., {Delaboudini{\`e}re}, J.-P., {et~al.}
  1999, \apj, 515, 842

\bibitem[{{Battaglia} \& {Kontar}(2013)}]{2013ApJ...779..107B}
{Battaglia}, M., \& {Kontar}, E.~P. 2013, \apj, 779, 107

\bibitem[{{Battaglia} {et~al.}(2015){Battaglia}, {Motorina}, \&
  {Kontar}}]{2015ApJ...815...73B}
{Battaglia}, M., {Motorina}, G., \& {Kontar}, E.~P. 2015, \apj, 815, 73

\bibitem[{Benvenuto {et~al.}(2013)Benvenuto, Schwartz, Piana, \&
  Massone}]{benvenuto2013expectation}
Benvenuto, F., Schwartz, R., Piana, M., \& Massone, A.~M. 2013, Astronomy \&
  Astrophysics, 555, A61.
\newblock \url{https://doi.org/10.1051/0004-6361/201321295}

\bibitem[{{Bian} {et~al.}(2014){Bian}, {Emslie}, {Stackhouse}, \&
  {Kontar}}]{2014ApJ...796..142B}
{Bian}, N.~H., {Emslie}, A.~G., {Stackhouse}, D.~J., \& {Kontar}, E.~P. 2014,
  \apj, 796, 142

\bibitem[{{Brown}(1971)}]{1971SoPh...18..489B}
{Brown}, J.~C. 1971, \solphys, 18, 489

\bibitem[{{Brown} {et~al.}(2003){Brown}, {Emslie}, \&
  {Kontar}}]{2003ApJ...595L.115B}
{Brown}, J.~C., {Emslie}, A.~G., \& {Kontar}, E.~P. 2003, \apjl, 595, L115

\bibitem[{{Carmichael}(1964)}]{1964NASSP..50..451C}
{Carmichael}, H. 1964, NASA Special Publication, 50, 451

\bibitem[{{Caspi} {et~al.}(2014){Caspi}, {McTiernan}, \&
  {Warren}}]{2014ApJ...788L..31C}
{Caspi}, A., {McTiernan}, J.~M., \& {Warren}, H.~P. 2014, \apjl, 788, L31

\bibitem[{Duval-Poo {et~al.}(2017)Duval-Poo, Massone, \&
  Piana}]{duval2017compressed}
Duval-Poo, M.~A., Massone, A.~M., \& Piana, M. 2017, in Sampling Theory and
  Applications (SampTA), 2017 International Conference on, IEEE, 677--681.
\newblock \url{https://doi.org/10.1109/SAMPTA.2017.8024408}

\bibitem[{{Duval-Poo} {et~al.}(2018){Duval-Poo}, {Piana}, \&
  {Massone}}]{2018A&A...615A..59D}
{Duval-Poo}, M.~A., {Piana}, M., \& {Massone}, A.~M. 2018, \aap, 615, A59

\bibitem[{{Emslie} {et~al.}(2008){Emslie}, {Hurford}, {Kontar}, {Massone},
  {Piana}, {Prato}, \& {Xu}}]{2008AIPC.1039....3E}
{Emslie}, A.~G., {Hurford}, G.~J., {Kontar}, E.~P., {et~al.} 2008, in American
  Institute of Physics Conference Series, Vol. 1039, American Institute of
  Physics Conference Series, ed. G.~{Li}, Q.~{Hu}, O.~{Verkhoglyadova}, G.~P.
  {Zank}, R.~P. {Lin}, \& J.~{Luhmann}, 3--10

\bibitem[{{Fleishman} {et~al.}(2018){Fleishman}, {Nita}, {Kuroda}, {Jia},
  {Tong}, {Wen}, \& {Zhizhuo}}]{2018arXiv180309847F}
{Fleishman}, G.~D., {Nita}, G.~M., {Kuroda}, N., {et~al.} 2018, ArXiv e-prints,
  arXiv:1803.09847

\bibitem[{Gallagher {et~al.}(2002)Gallagher, Dennis, Krucker, Schwartz, \&
  Tolbert}]{Gallagher2002}
Gallagher, P., Dennis, B., Krucker, S., Schwartz, R., \& Tolbert, A. 2002,
  Solar Physics, 210, doi:10.1023/A:1022422019779

\bibitem[{{Gallagher} {et~al.}(2002){Gallagher}, {Dennis}, {Krucker},
  {Schwartz}, \& {Tolbert}}]{2002SoPh..210..341G}
{Gallagher}, P.~T., {Dennis}, B.~R., {Krucker}, S., {Schwartz}, R.~A., \&
  {Tolbert}, A.~K. 2002, \solphys, 210, 341

\bibitem[{Guo {et~al.}(2012{\natexlab{a}})Guo, Emslie, Kontar, Benvenuto,
  Massone, \& Piana}]{Guo2012}
Guo, J., Emslie, A.~G., Kontar, E.~P., {et~al.} 2012{\natexlab{a}}, Astronomy
  {\&} Astrophysics, Volume 543, id.A53, 7 pp., 543, arXiv:1206.0477.
\newblock \url{http://arxiv.org/abs/1206.0477
  http://dx.doi.org/10.1051/0004-6361/201219341}

\bibitem[{Guo {et~al.}(2012{\natexlab{b}})Guo, Emslie, Massone, \&
  Piana}]{Guo2012a}
Guo, J., Emslie, A.~G., Massone, A.~M., \& Piana, M. 2012{\natexlab{b}}, The
  Astrophysical Journal, Volume 755, Issue 1, article id. 32, 6 pp. (2012).,
  755, arXiv:1206.2391.
\newblock \url{http://arxiv.org/abs/1206.2391
  http://dx.doi.org/10.1088/0004-637X/755/1/32}

\bibitem[{Guo {et~al.}(2013)Guo, Emslie, \& Piana}]{Guo2013}
Guo, J., Emslie, A.~G., \& Piana, M. 2013, The Astrophysical Journal, Volume
  766, Issue 1, article id. 28, 9 pp. (2013)., 766, arXiv:1303.1077.
\newblock \url{http://arxiv.org/abs/1303.1077
  http://dx.doi.org/10.1088/0004-637X/766/1/28}

\bibitem[{{Handy} {et~al.}(1999){Handy}, {Acton}, {Kankelborg}, {Wolfson},
  {Akin}, {Bruner}, {Caravalho}, {Catura}, {Chevalier}, {Duncan}, {Edwards},
  {Feinstein}, {Freeland}, {Friedlaender}, {Hoffmann}, {Hurlburt}, {Jurcevich},
  {Katz}, {Kelly}, {Lemen}, {Levay}, {Lindgren}, {Mathur}, {Meyer}, {Morrison},
  {Morrison}, {Nightingale}, {Pope}, {Rehse}, {Schrijver}, {Shine}, {Shing},
  {Strong}, {Tarbell}, {Title}, {Torgerson}, {Golub}, {Bookbinder}, {Caldwell},
  {Cheimets}, {Davis}, {Deluca}, {McMullen}, {Warren}, {Amato}, {Fisher},
  {Maldonado}, \& {Parkinson}}]{1999SoPh..187..229H}
{Handy}, B.~N., {Acton}, L.~W., {Kankelborg}, C.~C., {et~al.} 1999, \solphys,
  187, 229

\bibitem[{{Hirayama}(1974)}]{1974SoPh...34..323H}
{Hirayama}, T. 1974, \solphys, 34, 323

\bibitem[{{Holman}(2016)}]{2016JGRA..12111667H}
{Holman}, G.~D. 2016, Journal of Geophysical Research (Space Physics), 121, 11

\bibitem[{{Hurford} {et~al.}(2002){Hurford}, {Schmahl}, {Schwartz}, {Conway},
  {Aschwanden}, {Csillaghy}, {Dennis}, {Johns-Krull}, {Krucker}, {Lin},
  {McTiernan}, {Metcalf}, {Sato}, \& {Smith}}]{2002SoPh..210...61H}
{Hurford}, G.~J., {Schmahl}, E.~J., {Schwartz}, R.~A., {et~al.} 2002, \solphys,
  210, 61

\bibitem[{{Inglis} \& {Christe}(2014)}]{2014ApJ...789..116I}
{Inglis}, A.~R., \& {Christe}, S. 2014, \apj, 789, 116

\bibitem[{{Jeffrey} {et~al.}(2015){Jeffrey}, {Kontar}, \&
  {Dennis}}]{2015A&A...584A..89J}
{Jeffrey}, N.~L.~S., {Kontar}, E.~P., \& {Dennis}, B.~R. 2015, \aap, 584, A89

\bibitem[{{Kontar} {et~al.}(2011){Kontar}, {Hannah}, \&
  {Bian}}]{2011ApJ...730L..22K}
{Kontar}, E.~P., {Hannah}, I.~G., \& {Bian}, N.~H. 2011, \apjl, 730, L22

\bibitem[{{Kopp} \& {Pneuman}(1976)}]{1976SoPh...50...85K}
{Kopp}, R.~A., \& {Pneuman}, G.~W. 1976, \solphys, 50, 85

\bibitem[{{Krucker} {et~al.}(2010){Krucker}, {Hudson}, {Glesener}, {White},
  {Masuda}, {Wuelser}, \& {Lin}}]{2010ApJ...714.1108K}
{Krucker}, S., {Hudson}, H.~S., {Glesener}, L., {et~al.} 2010, \apj, 714, 1108

\bibitem[{{Krucker} {et~al.}(2008){Krucker}, {Hurford}, {MacKinnon}, {Shih}, \&
  {Lin}}]{2008ApJ...678L..63K}
{Krucker}, S., {Hurford}, G.~J., {MacKinnon}, A.~L., {Shih}, A.~Y., \& {Lin},
  R.~P. 2008, \apjl, 678, L63

\bibitem[{{Lemen} {et~al.}(2012){Lemen}, {Title}, {Akin}, {Boerner}, {Chou},
  {Drake}, {Duncan}, {Edwards}, {Friedlaender}, {Heyman}, {Hurlburt}, {Katz},
  {Kushner}, {Levay}, {Lindgren}, {Mathur}, {McFeaters}, {Mitchell}, {Rehse},
  {Schrijver}, {Springer}, {Stern}, {Tarbell}, {Wuelser}, {Wolfson}, {Yanari},
  {Bookbinder}, {Cheimets}, {Caldwell}, {Deluca}, {Gates}, {Golub}, {Park},
  {Podgorski}, {Bush}, {Scherrer}, {Gummin}, {Smith}, {Auker}, {Jerram},
  {Pool}, {Soufli}, {Windt}, {Beardsley}, {Clapp}, {Lang}, \&
  {Waltham}}]{2012SoPh..275...17L}
{Lemen}, J.~R., {Title}, A.~M., {Akin}, D.~J., {et~al.} 2012, \solphys, 275, 17

\bibitem[{{Lin} {et~al.}(2002){Lin}, {Dennis}, {Hurford}, {Smith}, {Zehnder},
  {Harvey}, {Curtis}, {Pankow}, {Turin}, {Bester}, {Csillaghy}, {Lewis},
  {Madden}, {van Beek}, {Appleby}, {Raudorf}, {McTiernan}, {Ramaty}, {Schmahl},
  {Schwartz}, {Krucker}, {Abiad}, {Quinn}, {Berg}, {Hashii}, {Sterling},
  {Jackson}, {Pratt}, {Campbell}, {Malone}, {Landis}, {Barrington-Leigh},
  {Slassi-Sennou}, {Cork}, {Clark}, {Amato}, {Orwig}, {Boyle}, {Banks},
  {Shirey}, {Tolbert}, {Zarro}, {Snow}, {Thomsen}, {Henneck}, {McHedlishvili},
  {Ming}, {Fivian}, {Jordan}, {Wanner}, {Crubb}, {Preble}, {Matranga}, {Benz},
  {Hudson}, {Canfield}, {Holman}, {Crannell}, {Kosugi}, {Emslie}, {Vilmer},
  {Brown}, {Johns-Krull}, {Aschwanden}, {Metcalf}, \&
  {Conway}}]{2002SoPh..210....3L}
{Lin}, R.~P., {Dennis}, B.~R., {Hurford}, G.~J., {et~al.} 2002, \solphys, 210,
  3

\bibitem[{{Pesnell} {et~al.}(2012){Pesnell}, {Thompson}, \&
  {Chamberlin}}]{2012SoPh..275....3P}
{Pesnell}, W.~D., {Thompson}, B.~J., \& {Chamberlin}, P.~C. 2012, \solphys,
  275, 3

\bibitem[{{Piana} {et~al.}(2007){Piana}, {Massone}, {Hurford}, {Prato},
  {Emslie}, {Kontar}, \& {Schwartz}}]{2007ApJ...665..846P}
{Piana}, M., {Massone}, A.~M., {Hurford}, G.~J., {et~al.} 2007, \apj, 665, 846

\bibitem[{{Priest} \& {Forbes}(2002)}]{2002A&ARv..10..313P}
{Priest}, E.~R., \& {Forbes}, T.~G. 2002, \aapr, 10, 313

\bibitem[{{Schmahl} {et~al.}(2007){Schmahl}, {Pernak}, {Hurford}, {Lee}, \&
  {Bong}}]{2007SoPh..240..241S}
{Schmahl}, E.~J., {Pernak}, R.~L., {Hurford}, G.~J., {Lee}, J., \& {Bong}, S.
  2007, \solphys, 240, 241

\bibitem[{{Schwartz} {et~al.}(2002){Schwartz}, {Csillaghy}, {Tolbert},
  {Hurford}, {McTiernan}, \& {Zarro}}]{2002SoPh..210..165S}
{Schwartz}, R.~A., {Csillaghy}, A., {Tolbert}, A.~K., {et~al.} 2002, \solphys,
  210, 165

\bibitem[{{Sim{\~o}es} \& {Kontar}(2013)}]{2013A&A...551A.135S}
{Sim{\~o}es}, P.~J.~A., \& {Kontar}, E.~P. 2013, \aap, 551, A135

\bibitem[{{Sturrock}(1966)}]{1966Natur.211..695S}
{Sturrock}, P.~A. 1966, \nat, 211, 695

\bibitem[{{Su} {et~al.}(2018){Su}, {Veronig}, {Hannah}, {Cheung}, {Dennis},
  {Holman}, {Gan}, \& {Li}}]{2018ApJ...856L..17S}
{Su}, Y., {Veronig}, A.~M., {Hannah}, I.~G., {et~al.} 2018, \apjl, 856, L17

\bibitem[{{Sui} \& {Holman}(2003)}]{2003ApJ...596L.251S}
{Sui}, L., \& {Holman}, G.~D. 2003, \apjl, 596, L251

\bibitem[{{Sui} {et~al.}(2004){Sui}, {Holman}, \&
  {Dennis}}]{2004ApJ...612..546S}
{Sui}, L., {Holman}, G.~D., \& {Dennis}, B.~R. 2004, \apj, 612, 546

\bibitem[{{Veronig} \& {Brown}(2004)}]{2004ApJ...603L.117V}
{Veronig}, A.~M., \& {Brown}, J.~C. 2004, \apjl, 603, L117

\bibitem[{{Veronig} {et~al.}(2006){Veronig}, {Karlick{\'y}}, {Vr{\v s}nak},
  {Temmer}, {Magdaleni{\'c}}, {Dennis}, {Otruba}, \&
  {P{\"o}tzi}}]{2006A&A...446..675V}
{Veronig}, A.~M., {Karlick{\'y}}, M., {Vr{\v s}nak}, B., {et~al.} 2006, \aap,
  446, 675

\bibitem[{{Woods} {et~al.}(2012){Woods}, {Eparvier}, {Hock}, {Jones},
  {Woodraska}, {Judge}, {Didkovsky}, {Lean}, {Mariska}, {Warren}, {McMullin},
  {Chamberlin}, {Berthiaume}, {Bailey}, {Fuller-Rowell}, {Sojka}, {Tobiska}, \&
  {Viereck}}]{2012SoPh..275..115W}
{Woods}, T.~N., {Eparvier}, F.~G., {Hock}, R., {et~al.} 2012, \solphys, 275,
  115

\bibitem[{{Xu} {et~al.}(2008){Xu}, {Emslie}, \&
  {Hurford}}]{2008ApJ...673..576X}
{Xu}, Y., {Emslie}, A.~G., \& {Hurford}, G.~J. 2008, \apj, 673, 576

\end{thebibliography}

\allauthors


\listofchanges

\end{document}